\documentclass[
preprint,
aps,
prapplied,
longbibliography,
superscriptaddress,
floatfix
]{revtex4-2}

\usepackage{amssymb}
\usepackage{amsmath,amsfonts}
\usepackage{array}
\usepackage{bm}
\usepackage{color}
\usepackage{multirow}

\usepackage[
    colorlinks=false,
    pdfborder={0 0 1},
    citebordercolor={0 0 1},
    linkbordercolor={1 1 1},
    urlbordercolor={0 0 1}
]{hyperref}

\usepackage{orcidlink}
\usepackage{adjustbox}
\usepackage[
    caption=false,
    labelformat=parens,
    labelfont=normalfont
]{subfig}

\setcitestyle{super,sort&compress}

\let\oldcite\cite
\renewcommand{\cite}[1]{{\color{blue}\oldcite{#1}}}

\begin{document}
%\linenumbers

\title{Robust Non-Adiabatic Holonomic Gating in Qutrits via Inverse-Engineered Pulse Shaping and Error Compensation}

\author{Jie Lu\orcidlink{0000-0002-6281-3504}}
\thanks{These authors contributed equally to this work.}
\email{lujie@shu.edu.cn}     
\affiliation{Department of Physics, Shanghai University, 200444 Shanghai, China}
\affiliation{Institute for Quantum Science and Technology, Shanghai University, 200444 Shanghai, China}

\author{Ji-Ze Han\orcidlink{0000-0002-6421-1523}$^*$}
%\thanks{These authors contributed equally to this work.}
\affiliation{China Mobile (Suzhou) Software Technology Co., Ltd., Suzhou 215163, China}
    
\author{Jie-Dong Huang}
\affiliation{Department of Physics, Shanghai University, 200444 Shanghai, China}

\author{Yang Qian}
\affiliation{Department of Physics, Shanghai University, 200444 Shanghai, China}

\author{Ying Yan\orcidlink{0000-0002-4898-6791}}
\email{yingyan@suda.edu.cn}  %% Mark Ying Yan as corresponding author
\affiliation{School of Optoelectronic Science and Engineering \textup{\&}
    Collaborative Innovation Center of Suzhou Nano Science and Technology, 
    Soochow University, Suzhou 215006, China}
\affiliation{Key Lab of Advanced Optical Manufacturing Technologies of Jiangsu Province \\
\textup{\&} Jiangsu Key Laboratory of Flexible Optoelectronics and Micro-Nano Manufacturing \\ 
Key Lab of Modern Optical Technology of Education Ministry of China, Soochow University, Suzhou 215006, China}

\author{Zhi-Guo Huang}
\email{huangzhiguo15@mails.ucas.ac.cn}
\affiliation{China Mobile (Suzhou) Software Technology Co., Ltd., Suzhou 215163, China}

\begin{abstract}

Systematic Rabi-amplitude and detuning errors remain important sources
of infidelity in high-fidelity quantum gates. We develop a robust
pulse-engineering scheme for non-adiabatic holonomic quantum computing
in a three-level $\Lambda$-type qutrit, combining inverse engineering
with time-dependent perturbative analysis. Pulse shaping eliminates the
leading second-order Rabi-amplitude contribution, while static detuning
introduces a distinct population-mediated channel that cannot be removed
within a single control loop. We therefore introduce a compensation loop
that exactly cancels the dominant second-order $O_{13}^{\delta}$
contribution, with the residual $O_{12}^{\delta}$ channel further
suppressed by pulse shaping.
Using the logical average gate fidelity over the complete computational
subspace, the optimized composite sequence reaches closed-system
fidelities of $99.88\%$--$99.99\%$ for four representative single-qubit
gates at $\epsilon=0.2$ and $\delta/2\pi=2$ MHz. With phenomenological
decoherence at $T_1=T_2=30~\mu{\rm s}$, the NOT and S gates retain
fidelities of $99.72\%$ and $99.79\%$, respectively, with a coherence-time
crossover near $0.58~\mu{\rm s}$.
These results identify the regime in which systematic-error suppression
outweighs the decoherence cost of the additional control loop.

\end{abstract}

\maketitle

\section{Introduction}
\label{sec:intro}
Quantum computing harnesses superposition and entanglement to
outperform classical hardware. While recent milestones in
superconducting processors, such as Google's `Willow'
chip~\cite{GoogleQuantumAIandCollaborators:2024efv}, mark significant progress in scaling and error correction, fault-tolerant operation remains constrained by residual physical errors and the substantial resource overhead required for quantum error correction
~\cite{google2023suppressing}. Although modern quantum processors are subject to complex noise environments including energy relaxation and stochastic dephasing, a substantial fraction of gate errors arises from systematic control imperfections---specifically Rabi frequency fluctuations and frequency detuning. These systematic errors continue to limit gate fidelities across diverse leading platforms, ranging from superconducting circuits~\cite{google2023suppressing, kim2023evidence} and trapped ions~\cite{moses2023race} to neutral atom systems~\cite{evered2023high}, necessitating robust control protocols that are resilient to hardware instabilities.

Geometric phases~\cite{berry1984quantal,aharonov1987phase}, defined by global properties of quantum evolution, provide inherent robustness against local perturbations~\cite{solinas2004robustness,thomas2011robustness}. Depending on the nature of the acquired geometric phases, quantum computation schemes based on these phases are systematically classified into two distinct frameworks: geometric quantum computation (GQC) and holonomic quantum computation (HQC)~\cite{zhang2023geometric}. Specifically, GQC utilizes Abelian geometric phases, which are scalar phase factors acquired by nondegenerate quantum states undergoing cyclic evolution. In contrast, HQC relies on non-Abelian geometric phases, which are matrix-valued quantum holonomies acquired by states within a multidimensional degenerate subspace~\cite{zhang2023geometric, sjoqvist2015geometric}. Because the non-Abelian geometric phase takes a matrix form, its intrinsic noncommutativity naturally facilitates the construction of a universal set of quantum gates. These ideas have also motivated fault-tolerant and accelerated holonomic constructions, including holonomic surface-code schemes and pulse-modulated adiabatic holonomic gates~\cite{zhang2018holonomic,zhang2025speeding}.

While early geometric gates used slow adiabatic evolution~\cite{zanardi1999holonomic,pachos1999non,pachos2001quantum,jones2000geometric,duan2001geometric}, their long operation times hindered high fidelity. In parallel, shortcut-to-adiabaticity and robust shaped-pulse techniques have provided general tools for accelerating adiabatic dynamics and suppressing control-induced errors~\cite{guery2019shortcuts,ruschhaupt2012optimally,daems2013robust}. Non-adiabatic geometric quantum computing (NGQC)~\cite{zhu2002implementation} and non-adiabatic holonomic quantum computing (NHQC)~\cite{sjoqvist2012non,sjoqvist2015geometric,sjoqvist2016conceptual,JunJing2025} leverage fast geometric phases to achieve high-fidelity gates, validated in systems like cold atoms and quantum dots. Recent developments have further extended geometric and holonomic gate operations to Rydberg atoms, superconducting circuits, defect-center systems, and bosonic encodings~\cite{liu2022optimized,xiao2024effective,su2024nonadiabatic,yun2024one,yun2024quantum,kang2022error}. Our present work strictly follows the NHQC framework. We encode the logical qubit within a two-dimensional computational subspace of a qutrit system, where the quantum operations are directly implemented via matrix-valued non-Abelian geometric phases generated through non-adiabatic cyclic evolution. Despite these advances, standard NHQC schemes remain susceptible to the aforementioned systematic errors in realistic experimental environments.

Single-qubit gate errors arise from both decoherence and systematic control imperfections. While NGQC and NHQC with short operation times naturally mitigate decoherence~\cite{carlini2013time}, systematic errors, such as Rabi and detuning errors, present substantial challenges that geometric protection alone cannot fully address~\cite{zheng2016comparison,colmenar2022conditions}. Accordingly, systematic-error-tolerant multiqubit holonomic gates, Rabi- and blockade-error-resilient all-geometric Rydberg gates, and time-optimal Rydberg NHQC protocols have been proposed to improve the practical robustness of geometric gate operations~\cite{wu2021systematic,su2023rabi,song2024fast}. To systematically analyze and overcome the remaining vulnerabilities in such operations, time-dependent perturbation theory can elucidate the explicit relationship between pulse engineering and gate fidelity, offering a pathway to higher robustness. Concurrently, to proactively design the requisite robust control fields, inverse engineering has emerged as a powerful theoretical tool. First applied to HQC in 2015 to rigorously formulate Orange-Slice-Shaped (OSS) path evolution~\cite{zhang2015fast}, this methodology has since been comprehensively generalized to construct fault-tolerant gates across broader contexts~\cite{zhang2023geometric}. Notably, related inverse-engineering and optimal-control approaches have successfully advanced geometric computation in Rydberg atoms and bosonic codes, enabling F{\"o}rster-resonance-based NHQC, invariant-based reverse engineering, and error-resistant binomial-code gates~\cite{liu2022optimized,xiao2024effective,kang2022error}.

Building upon these foundations, we develop a robust qutrit-specific NHQC scheme by combining inverse engineering with time-dependent perturbation theory.  We use time-dependent perturbation theory to derive analytic expressions for the fidelity of a quantum gate driven by a Hamiltonian of the form
\begin{equation}
H(t) = H_c(t) + H^{'}(t),\label{eq:Hc+H^{'}}
\end{equation}
where $H_c(t)$ is the ideal control Hamiltonian, and $H^{'}(t)$ represents the error Hamiltonian itself. This means that in the total Hamiltonian $H(t)$, the actual Rabi frequency and detuning are modified as $\Omega(t) \to (1+\epsilon)\Omega(t)$ and $\Delta(t) \to \Delta(t) + \delta$, respectively. It is worth noting that we parameterize $\epsilon$ as a dimensionless relative error, while $\delta$ is treated as an absolute frequency shift. This distinction captures their fundamentally different physical origins. Taking superconducting circuits as a representative example, amplitude errors typically arise from microwave electronics (e.g., amplifier fluctuations) and scale proportionally with the drive strength; conversely, detuning errors are dominated by intrinsic qubit frequency drifts ($\delta\omega_q$) induced by low-frequency $1/f$ flux noise. Since this ambient flux noise exists independently of the applied microwave power, describing $\delta$ as a quasi-static absolute constant decoupled from $\Omega(t)$ is physically more rigorous. Furthermore, while our general theoretical framework accommodates a time-dependent $\Delta(t)$, our specific holonomic protocol operates exactly on resonance ($\Delta(t) = 0$). Consequently, the actual detuning during the gate operation reduces purely to this static error $\delta$. Within this perturbative framework, we analytically demonstrate that the second-order Rabi error can be completely nullified through our tailored inverse-engineered pulse shaping. However, the second-order detuning error remains fundamentally tied to the excited state population, meaning it cannot be perfectly eliminated simultaneously by any single continuous pulse.

Crucially, to overcome this limitation, we introduce a composite
scheme incorporating a compensation pulse. Rather than eliminating
all detuning errors, the second loop produces an echo-like sign
reversal that exactly cancels the dominant population-mediated
$O_{13}^{\delta}$ contribution to second order, while pulse shaping
strongly suppresses the residual $O_{12}^{\delta}$ channel.
Although the compensation loop doubles the total gate duration,
our open-system analysis quantitatively identifies the regime in
which systematic-error suppression outweighs the additional
decoherence cost.

The paper is organized as follows: Section~\ref{sec:NHQC} introduces the physical model and the principles of NHQC in qutrit systems. Section~\ref{sec:perturbation_theory} presents our perturbation-based fidelity analysis and details the pulse engineering strategy for error mitigation. Section~\ref{sec:Numerical_simulation} validates our findings through comprehensive numerical simulations, comparing our robust pulses against standard schemes. Finally, Section~\ref{sec:conclusion} discusses the implications for scalable quantum computing and concludes the work.

 \section{Theoretical Framework and Physical Model}
\label{sec:NHQC}

\subsection{Principles of Non-Adiabatic Geometric Evolution}

To implement non-adiabatic quantum evolution, we construct a set of time-dependent orthonormal basis states $\{|\phi_n(t)\rangle\}$~\cite{Berry2009TransitionlessQD}, which satisfy the time-dependent Schr\"{o}dinger equation (with $\hbar=1$):
\begin{equation}
i\frac{d}{dt}|\phi_n(t)\rangle=H(t)|\phi_n(t)\rangle.
\label{eq:sch_eq}
\end{equation}
These states span the entire $N$-dimensional Hilbert space $\mathcal{H}$ of the system. Because each basis vector is an exact solution to Eq.~(\ref{eq:sch_eq}), their evolutions are dynamically decoupled. This ensures that the evolution is strictly unitary and no transitions occur between these distinct orthonormal basis states during the dynamics. When $|\phi_n(t)\rangle$ coincides with the instantaneous eigenstates of $H(t)$, the evolution reduces to the adiabatic limit.

 Despite the complex intermediate dynamics, we consider a cyclic evolution  of the system's basis states $|\phi_n(t)\rangle$,  meaning the basis vectors return to their initial configurations up to a phase factor $e^{i\gamma_n(\tau)}$ after a duration $\tau$:
\begin{equation}
|\phi_n(\tau)\rangle = e^{i\gamma_n(\tau)} |\phi_n(0)\rangle.
\label{eq:cyclic_phi}
\end{equation}
In this scenario, the  state projector undergoes a cyclic evolution, i.e., $\rho_n(\tau) = |\phi_n(\tau)\rangle\langle\phi_n(\tau)| = |\phi_n(0)\rangle\langle\phi_n(0)|$.

 We define a corresponding basis $\{|\chi_n(t)\rangle\}$ in the projective Hilbert space $\mathcal{P}(\mathcal{H})$ by removing the total phase:
\begin{equation}
|\chi_n(t)\rangle = e^{-i\gamma_n(t)} |\phi_n(t)\rangle.
\label{eq:chi_phi}
\end{equation}
The cyclic condition in Eq.~(\ref{eq:cyclic_phi}) implies that the projective state is also cyclic:
\begin{equation}
|\chi_n(\tau)\rangle = |\chi_n(0)\rangle = |\phi_n(0)\rangle.
\label{eq:cyclic_chi}
\end{equation}
The total phase $\gamma_n(\tau)$ acquired during the evolution can be decomposed into a dynamical phase $\gamma^d_n$ and a geometric phase $\gamma^g_n$:
\begin{align}
\gamma^d_n &= -\int_0^\tau \langle \phi_n(t)| H(t)| \phi_n(t)\rangle dt, \label{eq:gamma_d}\\
\gamma^g_n &= i \int_0^\tau \langle \chi_n(t)| \dot{\chi}_n(t)\rangle dt, \label{eq:gamma}
\end{align}
where $\gamma^g_n$ corresponds to the non-adiabatic Abelian geometric phase (Aharonov-Anandan phase)~\cite{aharonov1987phase}, which generalizes the Berry phase~\cite{berry1984quantal} to non-adiabatic evolutions.  By utilizing two different sets of basis vectors, we explicitly distinguish their physical origins: the dynamical phase is associated with the energy expectation in the full Hilbert space $\mathcal{H}$, whereas the geometric phase is determined by the connection in the projective Hilbert space $\mathcal{P}(\mathcal{H})$.

 To realize geometric quantum gates, the accumulated dynamical phase must be carefully managed. There are three primary strategies to achieve this:
\begin{enumerate}
\item Parallel Transport Condition: Enforcing the condition
\begin{equation}
\langle \phi_n(t)| H(t)| \phi_n(t)\rangle = 0
\label{eq:parallel_condition}
\end{equation}
at all times ensures that the dynamical phase vanishes identically ($\gamma_n^d = 0$). This is the standard condition for pure holonomic quantum computing.

\item Global Dynamical Phase Cancellation: Alternatively, ensuring that the time-integrated expectation value vanishes,
\begin{equation}
\int_0^\tau \langle \phi_n(t)| H(t)| \phi_n(t)\rangle dt = 0,
\label{eq:dynamical_cancellation}
\end{equation}
allows for the accumulation of instantaneous dynamical phases that cancel out by the end of the operation ($t=\tau$), a technique often employed in spin-echo sequences or specific pulse symmetry designs.

\item  Unconventional Geometric Gates: Instead of strict elimination, the dynamical phase can be engineered to be strictly proportional to the geometric phase ($\gamma_n^d \propto \gamma_n^g$). This approach, first pioneered as unconventional geometric quantum computation~\cite{zhu2003unconventional} and recently reviewed in Ref.~\cite{zhang2023geometric}, circumvents the stringent constraints of zero dynamical phase while still preserving the global geometric nature and robustness of the gate.
\end{enumerate}

Under the cyclic conditions (\ref{eq:cyclic_phi})--(\ref{eq:cyclic_chi}) and utilizing either of the dynamical phase elimination methods (\ref{eq:parallel_condition}) or (\ref{eq:dynamical_cancellation}), the time-evolution operator becomes purely geometric:
\begin{equation}
U(\tau)=\sum_n e^{i\gamma_n^g} |\phi_n(0)\rangle\langle \phi_n(0)|,
\label{eq:Utau_phi}
\end{equation}
 where the gate operation is characterized by the acquired geometric phases $\gamma_n^g$. As will be explicitly demonstrated in our specific qutrit model (Sec.~\ref{sec:three_level}), the exact cyclic excursion of the auxiliary states ($\alpha(0)=\alpha(\tau)=0$) guarantees that the system strictly returns to the computational subspace. This seamlessly translates the matrix-valued geometric phase into a target universal quantum gate.

\subsection{Holonomic Control in the Qutrit Subspace}
\label{sec:three_level}

We consider a qutrit system composed of three levels $|0\rangle$, $|1\rangle$, and $|e\rangle$, driven by pump and Stokes pulses with time-dependent frequencies $\nu_{S,P}(t)$. The pulse amplitudes are parameterized as 
\begin{align}
\Omega_P(t) &= \Omega(t) \sin(\theta/2) e^{-i\xi(t)}\\
\Omega_S(t) &= -\Omega(t) \cos(\theta/2) e^{i\phi} e^{-i\xi(t)}, 
\end{align}
where $\Omega(t)$ is the Rabi frequency envelope and $\phi$ is the relative phase.  Additionally, $\theta$ is the mixing angle that parameterizes the relative amplitude ratio between the two pulses (determining the rotation axis), and $\xi(t)$ is the time-dependent control phase introduced for inverse engineering.

Under the rotating wave approximation, the system Hamiltonian in the rotating frame is given by:
\begin{equation}
H^{(0)}(t) = \Delta(t) |e\rangle\langle e| + [\Omega_P(t) |0\rangle\langle e| + \Omega_S(t) |1\rangle\langle e| + \text{h.c.}],
\label{eq:3level_Hamiltonian}
\end{equation}
where the time-dependent detuning is defined as $\Delta(t) = 2\pi \nu_{S,P}(t) - \omega_{ei}$, with $\omega_{ei} = E_e - E_i$ ($i=0,1$).

\begin{figure}[htbp]
  \centering
  \includegraphics[width=\columnwidth]{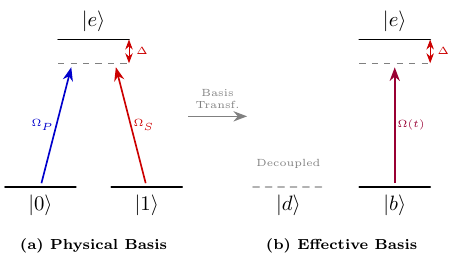}
 \caption{Schematic representation of the qutrit system dynamics. (a) The physical $\Lambda$-type configuration where pump ($\Omega_P$) and Stokes ($\Omega_S$) pulses drive transitions from the computational states $\{|0\rangle, |1\rangle\}$ to the auxiliary excited state $|e\rangle$ with time-dependent detuning $\Delta$. (b) The system in the transformed effective basis. Through the mapping defined by angles $\theta$ and $\phi$, the dark state $|d\rangle$ is decoupled from the evolution (indicated by the dashed line), reducing the dynamics to an effective two-level system involving only the bright state $|b\rangle$ and $|e\rangle$, driven by the effective Rabi frequency $\Omega(t)$.}
\label{fig:3level}
\end{figure}

The time-independent bare basis $\{|0\rangle, |1\rangle, |e\rangle\}$ spans the three-dimensional Hilbert space $\mathcal{H}$, with the qubit information encoded in the computational subspace $\{|0\rangle, |1\rangle\}$. The state vector $|\psi(t)\rangle$ evolves from an initial state $|\psi(0)\rangle$ within the computational subspace to a target state $|\psi_{\text{tg}}\rangle$, governed by an $SU(2)$ transformation:
\begin{equation}
|\psi(\tau)\rangle = U(\bm{n}, \gamma) |\psi(0)\rangle.
\end{equation}
To ensure cyclic evolution, the auxiliary excited state $|e\rangle$ must satisfy the condition of zero population at the boundaries $t=0$ and $t=\tau$.

Note that the Hamiltonian in Eq.~(\ref{eq:3level_Hamiltonian}) possesses a dark state $|d\rangle$ with a zero eigenvalue ($H(t)|d\rangle = 0$) and an orthogonal bright state $|b\rangle$, defined as:
\begin{align}
|d\rangle &= \cos\frac{\theta}{2} |0\rangle + \sin\frac{\theta}{2} e^{i\phi} |1\rangle, \label{eq:dark} \\
|b\rangle &= \sin\frac{\theta}{2} |0\rangle - \cos\frac{\theta}{2} e^{i\phi} |1\rangle. \label{eq:bright}
\end{align}
The subspace spanned by $\{|d\rangle, |b\rangle\}$ is thus equivalent to the computational subspace $\{|0\rangle, |1\rangle\}$.

Using these definitions, the Hamiltonian reduces to an effective two-level system involving only $|b\rangle$ and $|e\rangle$:
\begin{align}
H(t) = \Delta(t) |e\rangle\langle e| + [e^{-i\xi(t)} \Omega(t) |b\rangle\langle e| + \text{h.c.}], \label{eq:Hamiltonian_dbe}
\end{align}
where the dark state $|d\rangle$ is completely decoupled, as illustrated in Fig.~\ref{fig:3level}.

\begin{figure}[htbp]
  \centering
  \includegraphics[width=0.6\columnwidth]{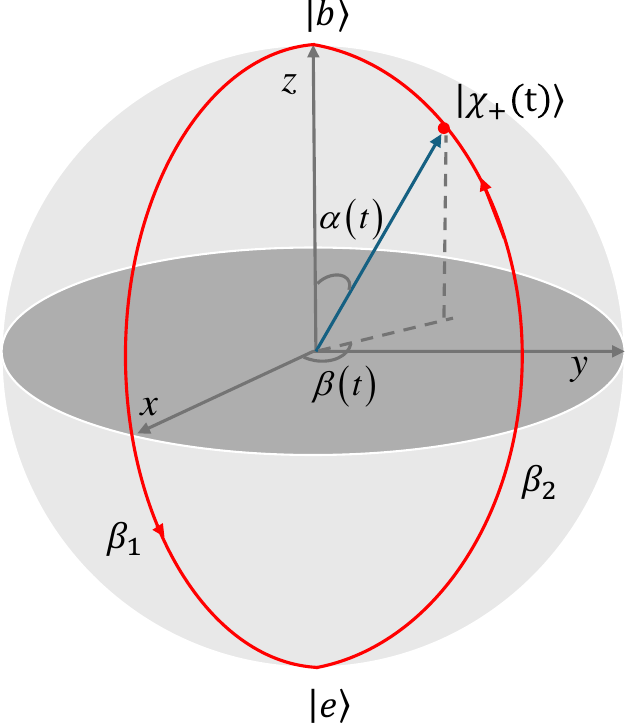}
  \caption{Geometric evolution path on the Bloch sphere for the OSS-NHQC scheme. The auxiliary state $|\chi_+(t)\rangle$ follows a longitudinal trajectory from the north pole ($|b\rangle$) to the south pole ($|e\rangle$) with constant azimuthal angle $\beta_1$ during the first half-period ($t \in [0, \tau/2]$), and returns to $|b\rangle$ with angle $\beta_2$ in the second half ($t \in (\tau/2, \tau]$). The solid angle enclosed by these two meridians determines the accumulated geometric phase $\gamma_g = \beta_1 - \beta_2$.}
\label{fig:BlochPath}
\end{figure}

We construct a time-dependent orthonormal basis $\{|\chi_+(t)\rangle, |\chi_-(t)\rangle\}$ as linear combinations of $|b\rangle$ and $|e\rangle$:
\begin{align}
|\chi_+(t)\rangle &= \cos\frac{\alpha(t)}{2} |b\rangle + \sin\frac{\alpha(t)}{2} e^{i\beta(t)} |e\rangle, \label{eq:chi+} \\
|\chi_-(t)\rangle &= \sin\frac{\alpha(t)}{2} |b\rangle - \cos\frac{\alpha(t)}{2} e^{i\beta(t)} |e\rangle, \label{eq:chi-}
\end{align}
where $\alpha(t)$ and $\beta(t)$ correspond to the polar and azimuthal angles on the Bloch sphere of the $\{|b\rangle, |e\rangle\}$ subspace, as depicted in Fig.~\ref{fig:BlochPath}. Together with $|d\rangle$, these states form the projective Hilbert space basis defined in Eq.~(\ref{eq:chi_phi}).
The cyclic evolution condition in Eq.~(\ref{eq:cyclic_chi}) requires the boundary conditions:
\begin{equation}
\alpha(0)=\alpha(\tau)=0.
\label{eq:cyclic_alpha}
\end{equation}
The cyclic condition here refers to the closure of the computational
subspace after one control loop; it does not require each logical state
to return to itself, since a nontrivial holonomic gate generally
implements a finite transformation within that subspace.

By substituting the physical pure states $\rho_\pm(t) = |\chi_\pm(t)\rangle\langle \chi_\pm(t)|$ into the von Neumann equation 
\begin{align}
i\frac{d\rho_\pm(t)}{dt}&=\left[H(t),\rho_\pm(t)\right],
\end{align}
we derive the control constraints:
\begin{align}
\Omega(t) &= -\frac{\dot{\alpha}(t)}{2\sin[\beta(t)-\xi(t)]}, \label{eq:Omegat} \\
\xi(t) &= \beta(t)-\arctan\left[\frac{\dot{\alpha}(t)}{\Delta(t)+\dot{\beta}(t)}\cot\alpha(t)\right]. \label{eq:xi}
\end{align}
These equations determine the driving pulse parameters $\Omega(t)$ and $\xi(t)$ based on the auxiliary functions $\alpha(t)$ and $\beta(t)$.

The resulting time-evolution operator is given by:
\begin{equation}
U(\tau)=|d\rangle\langle d|+e^{i\gamma}|b\rangle\langle b|,
\label{eq:Utau_db_gamma}
\end{equation}
which corresponds to an $SU(2)$ rotation in the computational subspace $\{|0\rangle, |1\rangle\}$:
\begin{equation}
U(\tau)=e^{i\frac{\gamma}{2}}e^{-i\frac{\gamma}{2}(\bm{n}\cdot \bm{\sigma})},
\label{eq:Utau_01_gamma}
\end{equation}
up to a global phase factor $e^{i\gamma/2}$. The rotation axis is $\bm{n} = (\sin\theta \cos\phi, \sin\theta \sin\phi, \cos\theta)$, where $\theta$ and $\phi$ are polar and azimuthal angles on the Bloch sphere of $\{|0\rangle, |1\rangle\}$.

The accumulated phase $\gamma$ may generally contain both geometric and dynamical components. A common approach to implement a holonomic gate is the Orange-Slice-Shaped (OSS) scheme, the concept of which was explicitly formulated in Ref.~\cite{zhang2015fast} (building upon related earlier ideas~\cite{li2002nonadiabatic}). It is defined by:
\begin{align}
\beta(t) &= \begin{cases}
\beta_1, & t \in [0, \tau/2], \\
\beta_2, & t \in (\tau/2, \tau].
\end{cases} \label{eq:beta_NHQC}
\end{align}
In this scheme, $|\chi_+(t)\rangle$ evolves from the north pole $|b\rangle$ to the south pole $|e\rangle$ with azimuthal angle $\beta_1$, and returns with $\beta_2$, as shown in Fig.~\ref{fig:BlochPath}. The accumulated geometric phase is:
\begin{align}
\gamma_g &= i \int_0^\tau \langle \chi_+(t) | \dot{\chi}_+(t) \rangle dt 
= -\int_0^\tau \dot{\beta}(t) \sin^2 \frac{\alpha(t)}{2} dt \nonumber \\
&= \beta_1 - \beta_2. \label{eq:geometric_phase}
\end{align}
while the dynamical phase in the resonance case ($\Delta(t) = 0$) is governed by:
\begin{align}
\gamma_d &= \frac{1}{2} \int_0^\tau \frac{\dot{\beta}(t) \sin^2 \alpha(t)}{\cos \alpha(t)} dt. \label{eq:dynamical_phase}
\end{align}
For the conventional OSS-NHQC scheme, $\beta(t)$ is constant within each segment ($\dot{\beta}(t) = 0$ for $t \neq \tau/2$).  The system thus follows a geodesic evolution where the dynamical phase naturally vanishes ($\gamma_d = 0$). However, the explicit integral in Eq.~(\ref{eq:dynamical_phase}) remains essential for our framework. To actively suppress systematic errors, we engineer generalized non-geodesic paths ($\dot{\beta}(t) \neq 0$) that rely on this exact integral to achieve global phase cancellation.

\section{Perturbative Error Analysis and Pulse Engineering}
\label{sec:perturbation_theory}

The analysis below builds on the time-dependent perturbative control
framework developed in Refs.~\cite{ruschhaupt2012optimally,daems2013robust}
and the inverse-engineering constructions in
Refs.~\cite{zhang2015fast,zhang2023geometric}.
Here we specialize these tools to the three-level $\Lambda$ qutrit
to identify and suppress the distinct Rabi-amplitude and detuning
error channels relevant to NHQC.

Consider a quantum system with state $|\psi(t)\rangle$ evolving under a Hamiltonian comprising a time-dependent control term $H_c(t)$ and a perturbation $H'(t)$. The Schr\"{o}dinger equation is:
\begin{align}
i\frac{d}{dt}|\psi(t)\rangle &= [H_c(t)+H'(t)]|\psi(t)\rangle.
\end{align}
In the absence of perturbation, we define the unperturbed state $|\psi^{(0)}(t)\rangle$ and the time-evolution operator $U^{(0)}(t)$ as:
\begin{align}
|\psi^{(0)}(t)\rangle &= U^{(0)}(t)|\psi(0)\rangle, \\
U^{(0)}(t) &= \mathcal{T} \exp\left[-i\int_0^t H_c(t')dt'\right].
\end{align}
To account for the time-dependent $H_c(t)$ and its instantaneous eigenstates, we introduce an error picture, which essentially corresponds to a toggling-frame transformation. This is distinct from the traditional interaction picture where the control Hamiltonian is typically time-independent. We define the state $|\psi_E(t)\rangle$ via:
\begin{align}
|\psi(t)\rangle &= U^{(0)}(t)|\psi_{E}(t)\rangle, \\
H_{E}'(t) &= U^{(0)\dagger}(t)H'(t)U^{(0)}(t).
\end{align}
In this error picture, $|\psi_E(t)\rangle$ satisfies the Schr\"{o}dinger equation:
\begin{align}
i\frac{d}{dt}|\psi_{E}(t)\rangle &= H_{E}'(t)|\psi_{E}(t)\rangle.
\end{align}
The states $|\psi(t)\rangle$ and $|\psi_E(t)\rangle$ are expanded in a perturbative series:
\begin{align}
|\psi(t)\rangle &= U(t)|\psi(0)\rangle \notag\\
&= |\psi^{(0)}(t)\rangle+|\psi^{(1)}(t)\rangle+|\psi^{(2)}(t)\rangle+\dots, \\
|\psi_{E}(t)\rangle &= U_E(t)|\psi(0)\rangle \notag\\
&= |\psi_{E}^{(0)}(t)\rangle+|\psi_{E}^{(1)}(t)\rangle+|\psi_{E}^{(2)}(t)\rangle+\dots,
\end{align}
where $U_E(t)$ is the error-picture time-evolution operator, expressed as a Dyson series:
\begin{align}
U_E(\tau) &= \mathcal{T}\exp\left[ -i\int^\tau_0 dt H'_E(t) \right] \nonumber\\
&= 1 - i\int_{0}^{\tau}dt_{1}H_{E}'(t_{1}) \nonumber\\
&\quad -\frac{1}{2}\int_{0}^{\tau}dt_{1}\int_{0}^{\tau}dt_{2}\mathcal{T}[H_{E}'(t_{1})H_{E}'(t_{2})] + \dots
\end{align}
Thus, the overlap amplitude between the evolved state $|\psi(\tau)\rangle$ and target state $|\psi_{\text{tg}}\rangle$ is:
\begin{align}
\langle \psi_{\text{tg}}|\psi(\tau)\rangle &= \left\langle \psi^{(0)}(0)\left|U_{E}(\tau)\right|\psi^{(0)}(0)\right\rangle \nonumber\\
&= 1 + O_1 + O_2 + \dots
\end{align}
with first- and second-order corrections:
\begin{align}
O_{1}(\tau) &= -i\int_{0}^{\tau}dt\left\langle \psi^{(0)}(t)\left|H'(t)\right|\psi^{(0)}(t)\right\rangle, \label{eq:O1tau}\\
O_2(\tau) &= -\frac{1}{2}\int_{0}^{\tau}dt_{1}\int_{0}^{\tau}dt_{2} \bigg[ E_{1}'(t_{1})E_{1}'(t_{2}) \nonumber\\
&\quad + \sum_{m\neq 1}H_{1m}'(t_{1})H_{m1}'(t_{2}) \bigg], \label{eq:O2tau}
\end{align}
where the matrix elements are defined as:
\begin{equation}
H_{1m}'(t) = \left\langle \psi_{1}^{(0)}(t)\left|H'(t)\right|\psi_{m}^{(0)}(t)\right\rangle.
\end{equation}
The orthonormal basis inserted into the second-order amplitude is:
\begin{align}
\left\{ \left|\psi_{n}^{(0)}(t)\right\rangle = U^{(0)}(t,0)|\psi_{n}^{(0)}(0)\rangle, \ n=1,2,3\dots \right\},
\end{align}
where $|\psi_{1}^{(0)}(t)\rangle$ coincides with the unperturbed state $|\psi^{(0)}(t)\rangle$(see Appendix~\ref{app:derivation} for details). The fidelity is:
\begin{align}
P(\tau) &= |\langle \psi_{\text{tg}}|\psi(\tau)\rangle|^2 = |1+O_1+O_2+\dots|^2 \nonumber\\
&\simeq 1 + O_{1}O_{1}^{*} + O_{2} + O_{2}^{*},
\label{eq:Ptau_general}
\end{align}
where the first-order correction vanishes as $O_1$ is purely imaginary, making the second-order term $\mathcal{O}(H^{'2}(t))$ the leading correction.  It is important to note that the validity of truncating the Dyson series at the second order relies on the smallness of the time-integrated perturbation strength, i.e., $\int_0^\tau ||H'(t)|| dt \ll 1$, rather than the absolute integration time $\tau$ itself. 
The second-order expansion is used here to identify the leading local
error channels and the corresponding pulse-design conditions near the
ideal operating point. At finite errors, where higher-order terms may
become appreciable, the gate performance is evaluated using the full
numerical evolution rather than relying on the truncated perturbative
expression. This distinction is explicitly tested in
Sec.~\ref{sec:Numerical_simulation}.

Beyond standard applications in two- and three-level quantum systems~\cite{ruschhaupt2012optimally,daems2013robust,liu2019plug,Xuezhengyuan2024a,Xuezhengyuan2024b}, this integration of perturbation theory and inverse engineering has been successfully generalized to suppress systematic errors in complex many-body platforms~\cite{kang2020heralded,liu2022optimized,xiao2024effective} and continuous-variable bosonic systems~\cite{chen2021fast,kang2022error}. Building upon this universality, we adapt this rigorous framework specifically to the qutrit configuration, offering a clear theoretical guide for designing highly robust nonadiabatic holonomic gates.

\subsection{Second-Order Fidelity Corrections under Systematic Errors}
\label{sec:fidelity_correction}

In leading quantum computing platforms—such as superconducting transmons, trapped ions, and neutral atoms—experimental operations are susceptible to two primary systematic errors: Rabi frequency fluctuations and frequency detuning. The Rabi error, modeled as $\Omega(t) \to (1 + \epsilon)\Omega(t)$, typically arises from spatial fluctuations in laser intensity or control electronics instability. The detuning error, $\Delta(t) \to \Delta(t) + \delta$, often stems from inhomogeneous broadening or magnetic field drifts in off-resonant systems. These systematic errors deviate the system's evolution from its intended holonomic trajectory, thereby reducing gate fidelity.

For the three-level system described in Section~\ref{sec:three_level}, we define the perturbation Hamiltonian $H^{'}(t)$ as:
\begin{equation}
H^{'}(t)=\delta|e\rangle\langle e|
+\epsilon\Omega(t)
\left[e^{-i\xi(t)}|b\rangle\langle e|+\mathrm{h.c.}\right].
\label{eq:H_prime_3level}
\end{equation}
Utilizing the complete orthonormal basis defined in Appendix~\ref{app:derivation} [Eqs.~(\ref{eq:complete_basisA1})--(\ref{eq:complete_basisA3})] and applying the general second-order fidelity formula [Eq.~(\ref{eq:Ptau_general})], we derive the fidelity at $t=\tau$:
\begin{align}
P(\tau) &\simeq 1 - \bigg[ \frac{1}{4} \left| c_b(0) (\delta O_{12}^{\delta} + \epsilon O_{12}^{\epsilon}) \right|^2 \nonumber \\
&\quad + \left| c_d^*(0) c_b^*(0) (\delta O_{13}^{\delta} + \epsilon O_{13}^{\epsilon}) \right|^2 \bigg],
\label{eq:Ptau_error_expansion}
\end{align}
where $c_b(0)$ and $c_d(0)$ are the probability amplitudes of the initial state $|\psi(0)\rangle = c_d(0) |d\rangle + c_b(0) |b\rangle$ in the $\{|d\rangle, |b\rangle\}$ subspace. The error coefficients are defined as:
\begin{align}
O_{12}^{\delta} &= \int_{0}^{\tau} dt \, e^{-i[\varphi_{+}(t)-\varphi_{-}(t)]} \sin\alpha(t), \label{eq:O12_delta}\\
O_{12}^{\epsilon} &= -\int_{0}^{\tau} dt \, e^{-i[\varphi_{+}(t)-\varphi_{-}(t)]} [\dot{\beta}(t)\sin\alpha(t) + i\dot{\alpha}(t)], \label{eq:O12_epsilon}\\
O_{13}^{\delta} &= \int_{0}^{\tau} dt \sin^2\frac{\alpha(t)}{2}, \label{eq:O13_delta}\\
O_{13}^{\epsilon} &= -\frac{1}{2} \int_{0}^{\tau} dt \, \dot{\beta}(t) \sin\alpha(t) \tan\alpha(t). \label{eq:O13_epsilon}
\end{align}
Here, $O_{12}^{\delta,\epsilon}$ characterizes the off-diagonal error
channel involving the bright--excited dynamics, whereas
$O_{13}^{\delta,\epsilon}$ describes the population-mediated channel
associated with the accumulated occupation of the auxiliary excited
state $|e\rangle$.

\subsection{Suppression of Rabi Errors via Pulse Shaping}
\label{sec:rabi_suppression}

To mitigate systematic errors while maintaining the geometric character of the operation, we propose a parameterized pulse design defined by:
\begin{align}
\beta(t)=\begin{cases}
a\sin\alpha(t)+\beta_{1}, & t\in\left[0,\frac{\tau}{2}\right],\\
b\sin\alpha(t)+\beta_{2}, & t\in\left(\frac{\tau}{2},\tau\right],
\end{cases}
\label{eq:beta_alpha}
\end{align}
where $a$ and $b$ are real parameters. The functional dependence of $\beta$ on $\alpha$ in Eq.~(\ref{eq:beta_alpha}) follows the parametrization strategy introduced in Ref.~\cite{xyyang2023}, which was originally designed to eliminate potential singularities in the Rabi frequency $\Omega(t)$.

Substituting Eq.~(\ref{eq:beta_alpha}) into the dynamical phase expression [Eq.~(\ref{eq:dynamical_phase})], we find that the contributions from the first and second halves of the evolution cancel each other out, provided that $a = b$. Consequently, the total phase is purely geometric with $\gamma = \gamma_g = \beta_1 - \beta_2$, preserving the holonomic nature of the gate for any $a=b$.

This general form reduces to the standard OSS-NHQC scheme~\cite{li2002nonadiabatic} for $a = b = 0$ and to the scheme in Ref.~\cite{xyyang2023} for $a = b = 1$; however, both specific cases exhibit significant residual error contributions. By further imposing the generalized condition:
\begin{equation}
a = b = 2n, \quad (n \in \mathbb{Z}\ \&\ n \neq 0, \pm 1), \label{eq:ab2n}
\end{equation}
we can simultaneously eliminate the second-order Rabi error terms, i.e., $O_{12}^\epsilon = O_{13}^\epsilon = 0$. In this case, the gate fidelity is dominated by detuning errors:
\begin{align}
P(\tau) \simeq 1 - \delta^2 \left[ \frac{1}{4} |c_b(0)|^2 |O_{12}^\delta|^2 + |c_d(0) c_b(0)|^2 |O_{13}^\delta|^2 \right].
\end{align}

The condition in Eq.~(\ref{eq:ab2n}) also helps mitigate these detuning errors. Assuming $\alpha(t)$ is symmetric about $t = \tau/2$, the error term $O_{12}^\delta$ can be expressed as $O_{12}^\delta = W (1 + e^{-i \gamma})$, where the integral factor is $W = \int_0^{\tau/2} e^{-i 2n \alpha(t)} \sin \alpha(t) dt$. The fidelity expression simplifies to:
\begin{align}
P(\tau) \simeq 1 - \delta^2 |c_b(0)|^2 \left[ \frac{|W|^2}{2} (1 + \cos \gamma) + |c_d(0)|^2 |O_{13}^\delta|^2 \right].
\end{align}
For $n = 2$ (i.e., $a=b=4$), the term $W$ is significantly suppressed by the rapidly oscillating factor $e^{-i 2n \alpha(t)}$, making it roughly an order of magnitude smaller than $O_{13}^\delta$. Thus, $O_{13}^\delta$ becomes the primary source of infidelity.

Physically, the term $O_{13}^\delta$ is directly related to the accumulated population in the excited state $|e\rangle$ during the evolution:
\begin{align}
\int^\tau_0 |c^{(0)}_e(t)|^2 dt &= |c_b(0)|^2 \int^\tau_0 \sin^2\frac{\alpha(t)}{2}dt \nonumber\\
&= |c_b(0)|^2 O_{13}^\delta.
\label{eq:O13delta_estate}
\end{align}
Although optimizing the pulse envelope $\alpha(t)$ can reduce this population, it remains strictly non-zero for any single-loop evolution.  Therefore, unlike the second-order Rabi error, which can be completely nullified through tailored pulse shaping alone, this implies that  the detuning error governed by $O_{13}^\delta$ cannot be fully eliminated within a single gate pulse. To address this limitation, we introduce a compensation pulse in the
next subsection to cancel the dominant $O_{13}^{\delta}$ contribution
through an echo-like interference mechanism.

\subsection{Cancellation of Detuning Errors via Compensation Pulses}
\label{sec:compensation_pulse}

Compensation pulses, employed in quantum optics~\cite{roos2004quantum},
resemble spin-echo techniques in canceling unwanted phase contributions.
We introduce a compensation pulse to exactly cancel the dominant
population-mediated second-order detuning channel
$O_{13}^{\delta}$ in the three-level NHQC system, while the residual
$O_{12}^{\delta}$ contribution is further suppressed by pulse shaping.

We apply a gate pulse over $t \in [0, \tau]$, followed by a compensation pulse over $t \in [\tau, 2\tau]$, modifying the Hamiltonian to $\tilde{H}(t) = \tilde{H}_c(t) + H^{'}(t)$, where $\tilde{H}_c(t)$ replaces $H_c(t)$ after $t = \tau$, while the perturbation $H^{'}(t)$ remains unchanged. We emphasize that this composite sequence is fundamentally different from a straightforward repetition of the gate pulse. Mathematically, a simple repetition would cause the error integrals to sum constructively, amplifying the second-order infidelity. In contrast, our scheme is designed to act as a geometric spin-echo. To realize the error cancellation, we configure the second stage by swapping the dark state $|d\rangle$ and bright state $|b\rangle$ via the parameter transformation $\theta \to \tilde{\theta} = \pi - \theta$ and $\phi \to \tilde{\phi} = \pi + \phi$. 
By setting $\tilde{\gamma}=0$, the second stage forms a cyclic compensation loop that implements the identity operation on the
logical computational subspace at the end of the sequence
~\cite{yan2019robust}.

The final state at $t = 2\tau$ is expanded in a perturbation series:
\begin{align}
|\tilde{\psi}(2\tau)\rangle &= \tilde{U}(2\tau, \tau) |\psi(\tau)\rangle = \tilde{U}(2\tau, \tau) U(\tau, 0) |\psi(0)\rangle \nonumber \\
&= |\tilde{\psi}^{(0)}(2\tau)\rangle + |\tilde{\psi}^{(1)}(2\tau)\rangle + |\tilde{\psi}^{(2)}(2\tau)\rangle + \cdots.
\end{align}
For the second stage, the same orthonormal basis is propagated
continuously according to
\begin{align}
\left\{
|\tilde{\psi}_n^{(0)}(t)\rangle
=
\tilde U^{(0)}(t,\tau)
|\psi_n^{(0)}(\tau)\rangle,\quad n=1,2,3
\right\}.
\end{align}
The fidelity, accounting for both evolution stages, is derived as (see Appendix~\ref{app:derivation} for details):
\begin{align}
P(2\tau) &\simeq 1-\sum_{m\neq1}\left\{ \left|\int_{0}^{\tau}dtH_{1m}^{'}(t)\right|^2 + \left|\int_{\tau}^{2\tau}dt\tilde{H}_{1m}^{'}(t)\right|^2\right.\nonumber\\
&\quad \left.+\int_{0}^{\tau}dt_{1}\int_{\tau}^{2\tau}dt_{2}\left[H_{m1}^{'}(t_{1})\tilde{H}_{1m}^{'}(t_{2})+\text{h.c.}\right]\right\}.
\label{eq:P2tau}
\end{align}

Since the second-order Rabi error is already eliminated under the
condition in Eq.~(\ref{eq:ab2n}), we focus on the detuning error,
$H^{'}=\delta|e\rangle\langle e|$. For this specific error type, the parameter transformation induces
the sign reversal
$\tilde H_{13}^{'}(t)=-H_{13}^{'}(t-\tau)$
(see Appendix~\ref{app:derivation}), which makes the
$m=3$ contribution in Eq.~(\ref{eq:P2tau}) cancel exactly.

Consequently, the residual second-order contribution originates from
the $m=2$ channel. For an arbitrary logical input state, the resulting
state fidelity can be written as
\begin{equation}
P_{\psi}(2\tau)
\simeq
1-\frac{\delta^2}{4}
\left[
|c_b|^2|\mathcal I_1|^2
+|c_d|^2|\mathcal I_2|^2
+2{\rm Re}\left(
c_d^*c_b\mathcal I_1^*\mathcal I_2
\right)
\right].
\label{eq:P2tau_3level}
\end{equation}
The first two terms describe the residual $O_{12}^{\delta}$ channel,
while the last term is an input-state-dependent interference term.
Its Haar average vanishes, as shown in
Appendix~\ref{app:haar_average_fidelity}.

The key result is therefore not the cancellation of all detuning
errors, but the exact cancellation of the dominant
population-mediated $O_{13}^{\delta}$ channel to second order.
The remaining $O_{12}^{\delta}$ contribution is strongly reduced by
the oscillatory integral $W = \int_0^{\tau/2} e^{-i 2n \alpha(t)} \sin \alpha(t) dt$ for the $a=b=4$ trajectory.

It is worth noting that the robustness of this scheme against imperfections in the compensation pulse itself relies on the assumption that the systematic errors remain quasi-static over the doubled gate duration. If the control fields introduce time-dependent high-frequency fluctuations, the cancellation efficiency will inevitably degrade.

\section{Numerical Validation and Robustness Analysis}
\label{sec:Numerical_simulation}

\subsection{Pulse Configuration and State Evolution}
\label{sec:pulse_config}

Practical implementation requires smooth control envelopes with limited
bandwidth. According to Eq.~(\ref{eq:Omegat}), the boundary condition
$\Omega(0)=\Omega(\tau)=0$ can be satisfied by imposing
$\dot{\alpha}(0)=\dot{\alpha}(\tau)=0$. We choose
\begin{equation}
\alpha(t)=\pi\sin^2\left(\pi\frac{t}{\tau}\right),
\label{alpha}
\end{equation}
with $\beta(t)$ defined by Eq.~(\ref{eq:beta_alpha}) using $n=2$.
For the compensation stage over $t\in[\tau,2\tau]$, $\alpha(t)$ follows
the same functional form, while the corresponding parameters are chosen
to realize the identity compensation loop with $\tilde{\gamma}=0$.

\begin{figure}[htbp]
\centering
\subfloat[]{
\includegraphics[width=0.48\columnwidth]{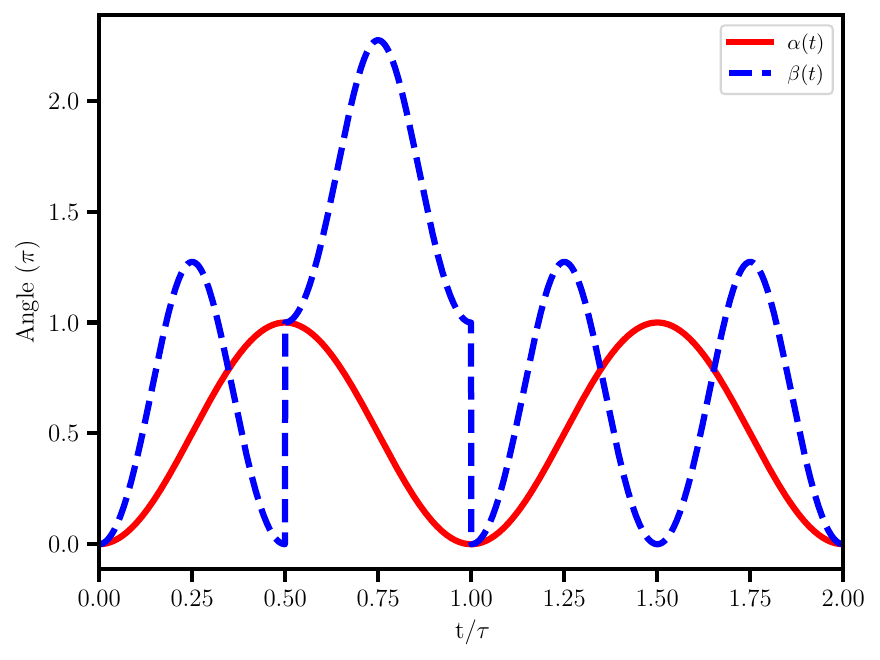}}
\hfill
\subfloat[]{
\includegraphics[width=0.48\columnwidth]{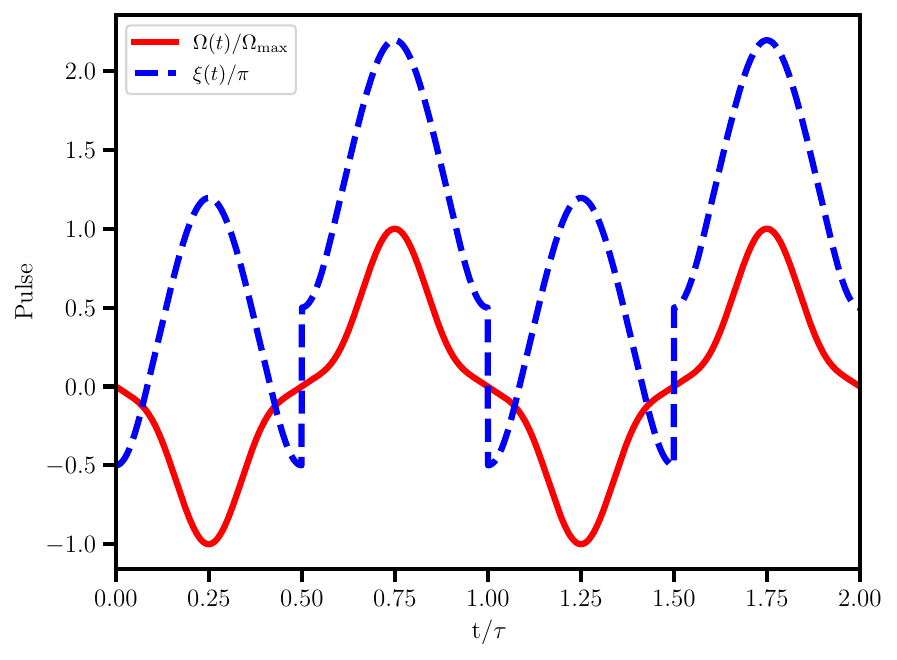}}
\caption{Time evolution of the control parameters for the robust NOT
gate ($a=b=4$). (a) Auxiliary angles $\alpha(t)/\pi$ and
$\beta(t)/\pi$. (b) Normalized Rabi envelope
$\Omega(t)/\Omega_{\rm max}$ and control phase $\xi(t)/\pi$.
The complete sequence contains the gate stage
$t\in[0,\tau]$ and the compensation stage $t\in[\tau,2\tau]$.}
\label{fig:pulse}
\end{figure}

Figure~\ref{fig:pulse} shows the resulting pulse profiles for the
robust NOT gate. The control envelopes remain continuous and smooth
throughout the complete sequence.

To illustrate the state dynamics, we first consider representative
logical input states written as
\begin{equation}
|\psi(0)\rangle
=
\cos\theta_0|0\rangle
+
\sin\theta_0e^{i\phi_0}|1\rangle .
\label{eq:psi0}
\end{equation}
Here $\theta_0$ and $\phi_0$ specify the input state, whereas
$\gamma$, $\theta$, and $\phi$ determine the target gate.
The representative input-output pairs used only to illustrate the
state dynamics are summarized in Table~\ref{Table: parameters}.

\begin{table}[htbp]
\centering
\caption{Parameters of the four representative quantum gates.
The parameters $\theta_0$ and $\phi_0$ specify the input state,
whereas $\gamma$, $\theta$, and $\phi$ determine the target gate.}
\label{Table: parameters}
\setlength{\tabcolsep}{1mm}{
\begin{adjustbox}{max width=\columnwidth}
{
\begin{tabular}{c c c c c c c c}
\hline
\hline
Gate & $\theta_0$ & $\phi_0$ & $\gamma$ & $\theta$ & $\phi$
& $|\psi(0)\rangle$ & $|\psi_{\rm tg}\rangle$ \\
\hline
NOT
& 0 & 0 & $\pi$ & $\frac{\pi}{2}$ & 0
& $|0\rangle$
& $|1\rangle$ \\
Hadamard
& 0 & 0 & $\pi$ & $\frac{\pi}{4}$ & 0
& $|0\rangle$
& $\frac{1}{\sqrt{2}}(|0\rangle+|1\rangle)$ \\
S
& $\frac{\pi}{4}$ & 0 & $\frac{\pi}{2}$ & 0 & 0
& $\frac{1}{\sqrt{2}}(|0\rangle+|1\rangle)$
& $\frac{1}{\sqrt{2}}(|0\rangle+i|1\rangle)$ \\
T
& $\frac{\pi}{4}$ & 0 & $\frac{\pi}{4}$ & 0 & 0
& $\frac{1}{\sqrt{2}}(|0\rangle+|1\rangle)$
& $\frac{1}{\sqrt{2}}(|0\rangle+e^{i\pi/4}|1\rangle)$ \\
\hline
\hline
\end{tabular}
}
\end{adjustbox}
}
\end{table}

\begin{figure*}[htp]
\centering
\subfloat[NOT gate]{
\includegraphics[width=0.45\textwidth,height=5cm]{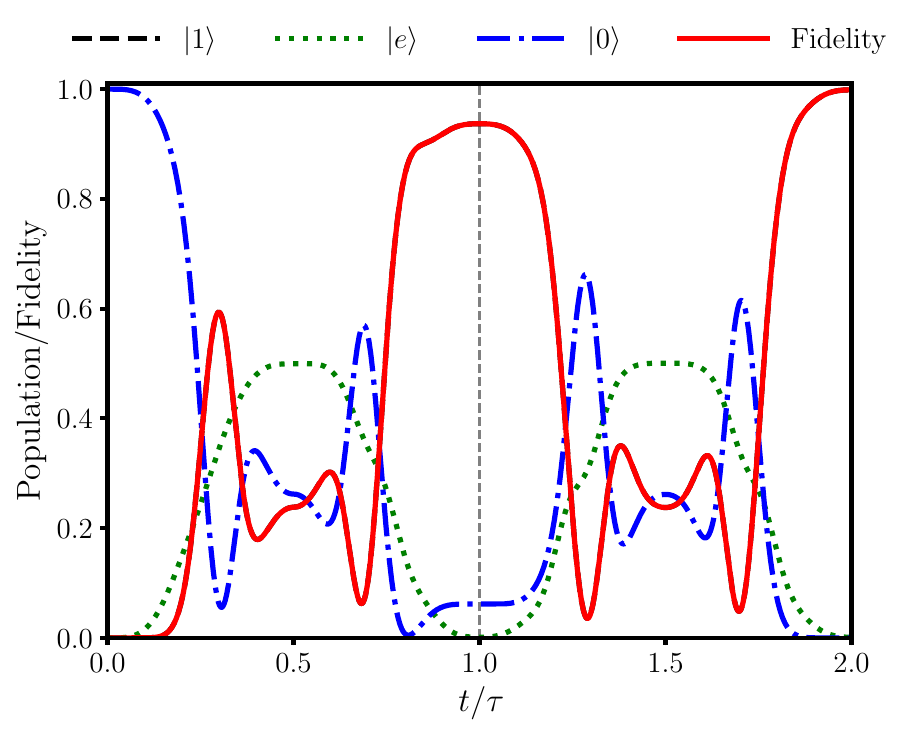}}
\hfill
\subfloat[Hadamard gate]{
\includegraphics[width=0.45\textwidth,height=5cm]{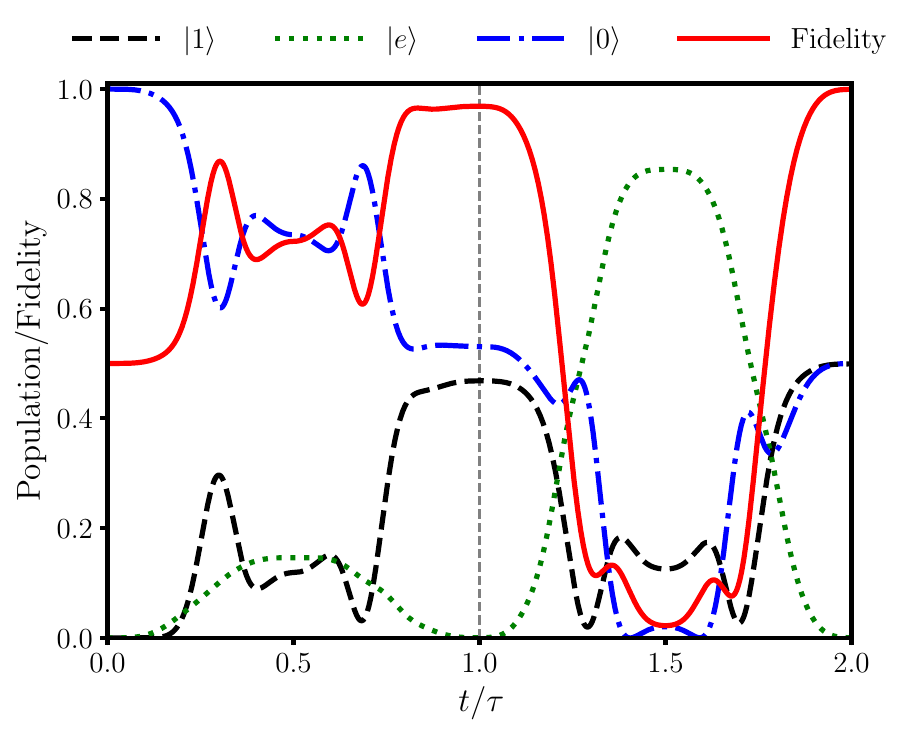}}
\\
\subfloat[S gate]{
\includegraphics[width=0.45\textwidth,height=5cm]{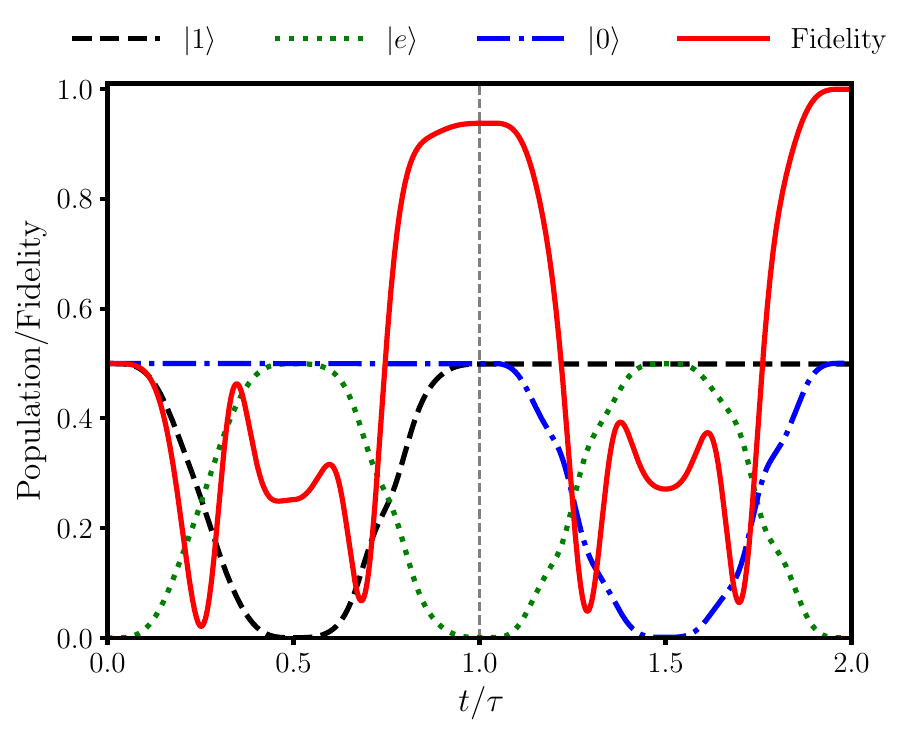}}
\hfill
\subfloat[T gate]{
\includegraphics[width=0.45\textwidth,height=5cm]{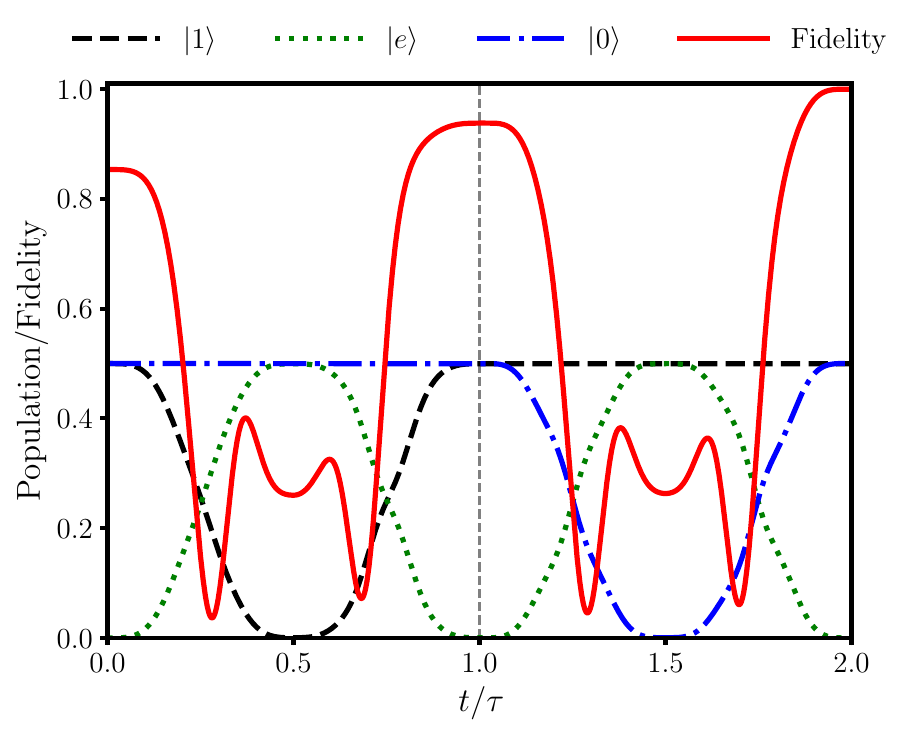}}
\caption{State-fidelity and population dynamics for representative
quantum gates under $\epsilon=0.2$ and $\delta/2\pi=2$ MHz.
The solid red curve denotes the state fidelity, while the blue dash-dotted,
black dashed, and green dotted curves denote the populations of
$|0\rangle$, $|1\rangle$, and $|e\rangle$, respectively. 
The grey dashed line separates the gate and compensation stages.}
\label{fig:population_NOT}
\end{figure*}

For systematic errors $\epsilon=0.2$ and
$\delta/2\pi=2$ MHz, Fig.~\ref{fig:population_NOT} shows the
state-fidelity and population dynamics over the complete composite
sequence. The first stage implements the target gate, while the second
stage realizes the error-compensation loop.

To characterize the gate performance over the complete computational
subspace rather than for a particular input state, we use the logical
average gate fidelity. For an evolution of total duration $T$, where
$T=\tau$ for a single-loop sequence and $T=2\tau$ when the compensation
pulse is included, we define
\begin{equation}
M=P_{\rm L}U(T)P_{\rm L},
\label{eq:logical_projected_operation}
\end{equation}
and evaluate
\begin{equation}
F_{\rm avg}
=
\frac{
{\rm Tr}(M^\dagger M)
+
\left|
{\rm Tr}(U_{\rm tg}^{\dagger}M)
\right|^2
}
{d(d+1)},
\qquad d=2.
\label{eq:Favg_definition}
\end{equation}

The numerical calculations evaluate this quantity using the six Pauli
eigenstates, whose equally weighted average is exactly equivalent to
the Haar average for the present single-qubit gate fidelity.
The corresponding Haar-average derivation and its relation to the
perturbative state fidelity are given in
Appendix~\ref{app:haar_average_fidelity}.

For the optimized $a=b=4$ pulse, we set the single-loop duration to
$\tau=80$ ns, corresponding to
$\Omega_{\rm max}/2\pi\simeq40.5$ MHz. The compensated sequence has a
total duration of $2\tau=160$ ns.

For the open-system simulations, we use
$T_1=T_2=30~\mu{\rm s}$ as a conservative coherence benchmark.
As an order-of-magnitude coherence reference, the 105-qubit Willow processor reported in
Ref.~\cite{google2025quantum} has mean coherence times of approximately
$T_1=68~\mu{\rm s}$ and $T_{2,\mathrm{CPMG}}=89~\mu{\rm s}$.
These values are used only as representative experimentally accessible
control scales.

\subsection{Robustness against Single-Parameter Errors}
\label{sec:single_error}

We first consider Rabi-amplitude and detuning errors separately.

\textbf{Suppression of Rabi Errors:}

\begin{figure*}[t]
\centering
\subfloat[NOT gate]{
\includegraphics[width=0.48\textwidth]{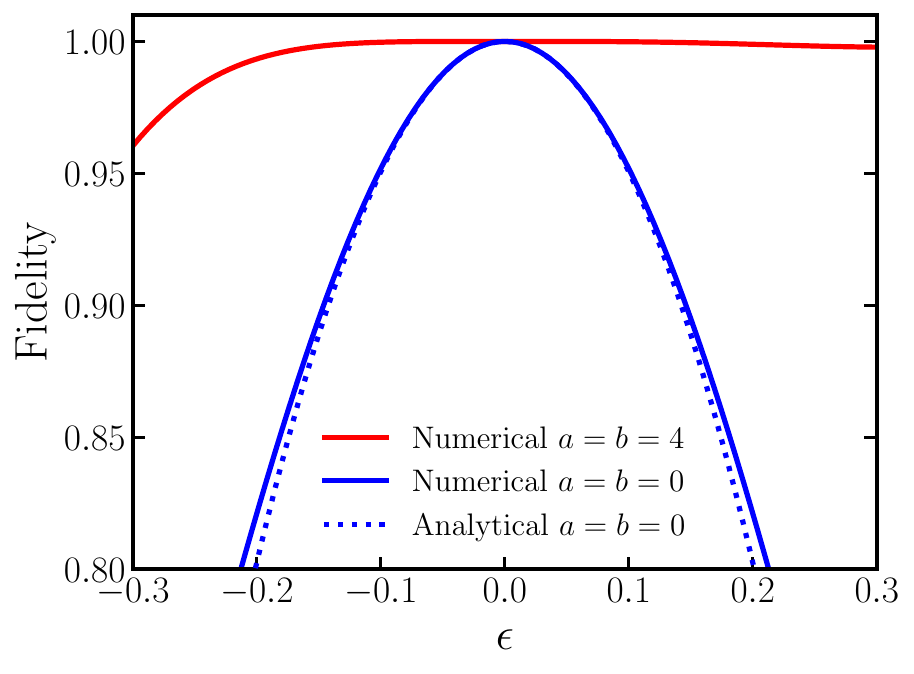}}
\hfill
\subfloat[Hadamard gate]{
\includegraphics[width=0.48\textwidth]{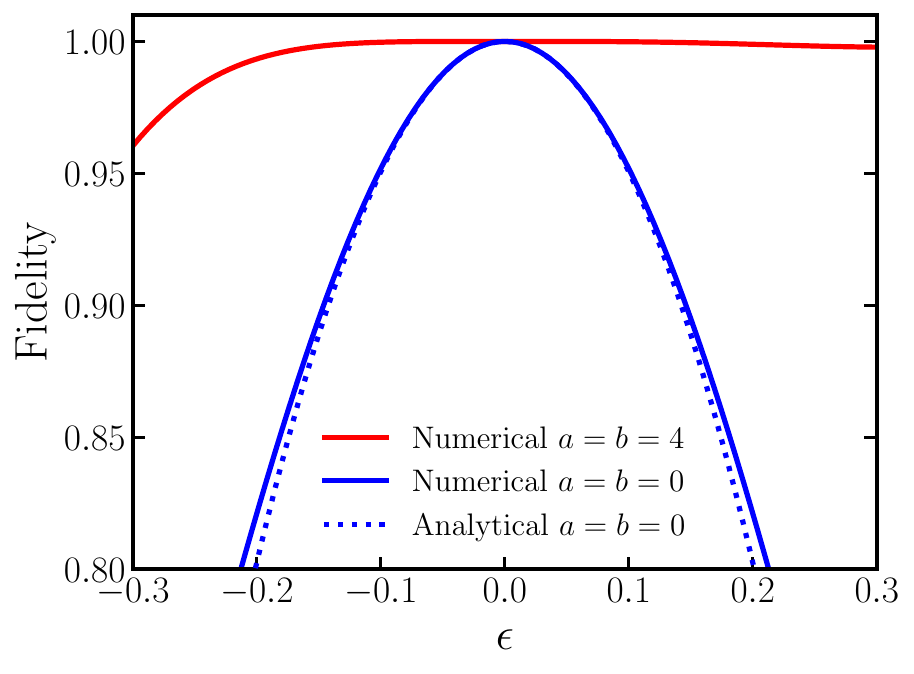}}
\\
\subfloat[S gate]{
\includegraphics[width=0.48\textwidth]{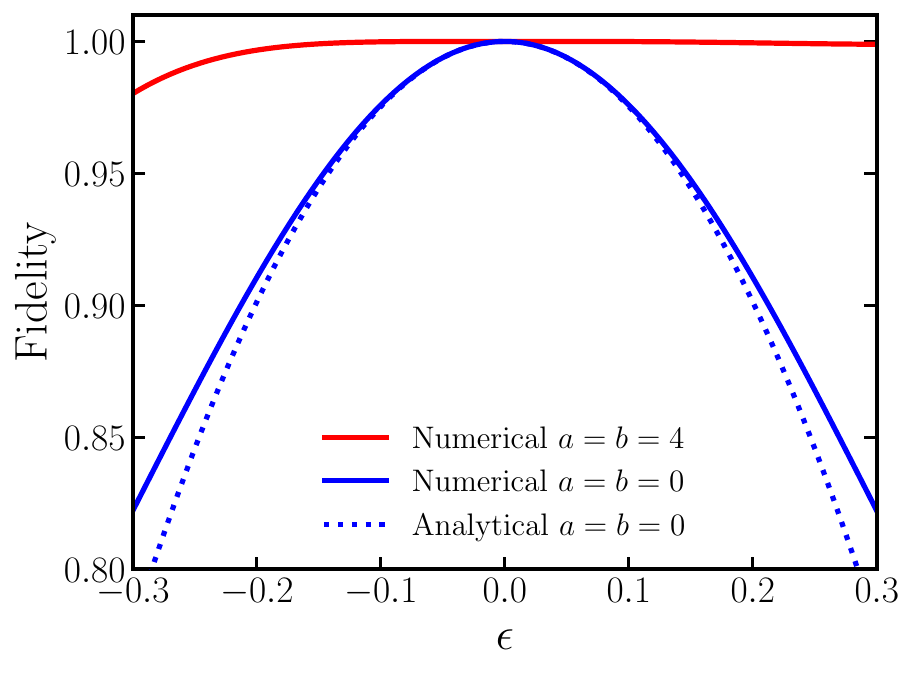}}
\hfill
\subfloat[T gate]{
\includegraphics[width=0.48\textwidth]{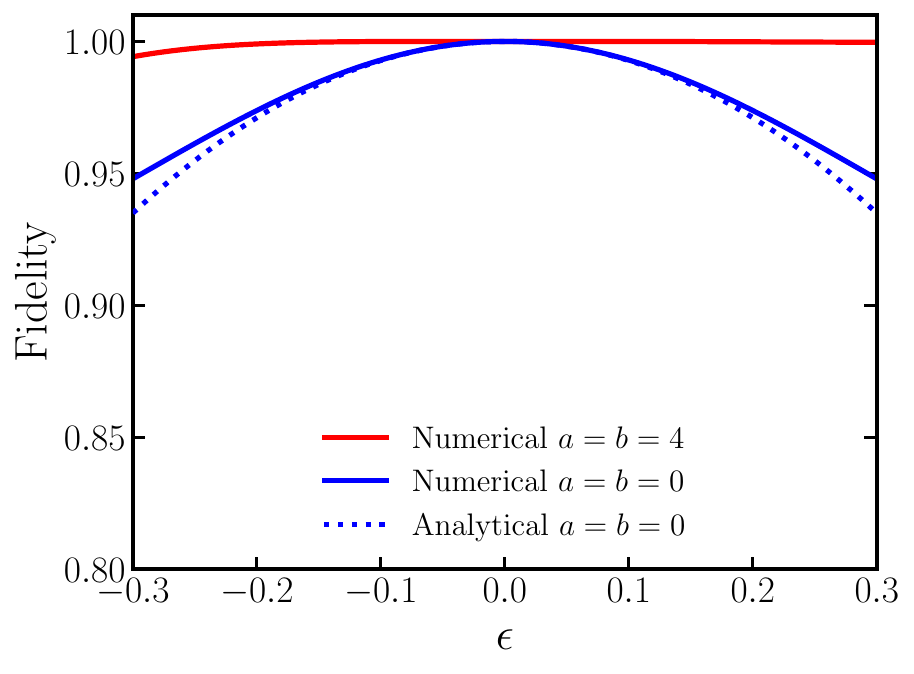}}
\caption{Logical average gate fidelity as a function of the
Rabi-amplitude error $\epsilon$.
The red solid curves show the numerical results for $a=b=4$, the blue
solid curves show the numerical results for $a=b=0$, and the blue
dotted curves show the corresponding second-order analytical results
for $a=b=0$.}
\label{fig:fidelity_epsilon}
\end{figure*}

For Rabi-amplitude errors, the $a=b=4$ pulse satisfies the condition
derived in Sec.~\ref{sec:perturbation_theory} and eliminates the
leading second-order Rabi-error contribution.
Figure~\ref{fig:fidelity_epsilon} compares this optimized sequence
with the conventional $a=b=0$ OSS-NHQC protocol.

For consistency with the numerical logical average gate fidelity, the
perturbative coefficients are averaged over the complete logical input
space. The corresponding Haar-averaged result is derived in
Appendix~\ref{app:haar_average_fidelity}, and the second-order
coefficients for the four gates are summarized in
Table~\ref{Table:infidelity_epsilon}.

\begin{table}[htbp]
\centering
\caption{Haar-averaged second-order Rabi-error coefficients
$f_{\mathrm{Infi}}$ for the conventional $a=b=0$ sequence, defined by
$1-F_{\mathrm{avg}}^{(2)}=\epsilon^2f_{\mathrm{Infi}}$.
}
\label{Table:infidelity_epsilon}
\setlength{\tabcolsep}{3mm}{
\begin{tabular}{c c c c c}
\hline
\hline
Gate & NOT & Hadamard & S & T \\
\hline
$f_{\mathrm{Infi}}$
& $\frac{1}{2}\pi^2$
& $\frac{1}{2}\pi^2$
& $\frac{1}{4}\pi^2$
& $\frac{1}{4}\pi^2\left(1-\frac{\sqrt{2}}{2}\right)$ \\[6pt]
\hline
\hline
\end{tabular}
}
\end{table}

The numerical results confirm the strong suppression of Rabi-amplitude
errors by the optimized pulse. For $a=b=4$, the quadratic contribution
vanishes and the remaining infidelity is governed by fourth- and
higher-order terms.

\textbf{Cancellation of Detuning Errors:}

For static detuning errors, we compare the conventional and optimized
trajectories both with and without the compensation pulse.
The NOT and S gates are used as representative examples.

For compact notation, we write
$O_{12}^{\delta}=A(1+e^{-i\gamma})$, where
$A=Q$ for the conventional $a=b=0$ sequence and $A=W$ for the
optimized $a=b=4$ sequence, with
$Q=\int_0^{\tau/2}\sin\alpha(t)dt$ and
$W=\int_0^{\tau/2}e^{-i4\alpha(t)}\sin\alpha(t)dt$.

The corresponding Haar-averaged second-order detuning fidelities,
with and without the compensation pulse, are derived in
Appendix~\ref{app:haar_average_fidelity}.
The resulting coefficients for the NOT and S gates are summarized in
Table~\ref{Table:infidelity_delta}.

The compensation loop reverses the sign of the
population-mediated $O_{13}^{\delta}$ error amplitude, so this dominant
contribution cancels between the two stages.
The residual $O_{12}^{\delta}$ channel remains, but is strongly
suppressed in the optimized trajectory through $|W|\ll Q$.
Thus, the compensation pulse cancels the dominant
$O_{13}^{\delta}$ contribution rather than all detuning errors, while
pulse shaping suppresses the remaining $O_{12}^{\delta}$ channel.

\begin{table}[htbp]
\centering
\caption{Haar-averaged second-order detuning-error coefficients
$f_{\mathrm{Infi}}$ defined by
$1-F_{\mathrm{avg}}^{(2)}=\delta^2f_{\mathrm{Infi}}$,
with and without the compensation pulse (CP).}
\label{Table:infidelity_delta}

\setlength{\tabcolsep}{3mm}
\begin{tabular}{c c c c c}
\hline
\hline
& \multicolumn{2}{c}{$a=b=0$}
& \multicolumn{2}{c}{$a=b=4$} \\
$f_{\mathrm{Infi}}$
& No CP & CP & No CP & CP \\
\hline
NOT\rule{0pt}{3.2ex}
& $\tfrac{1}{6}|O_{13}^{\delta}|^2$
& $\tfrac{1}{2}Q^2$
& $\tfrac{1}{6}|O_{13}^{\delta}|^2$
& $\tfrac{1}{2}|W|^2$ \\[8pt]
S
& $\tfrac{1}{6}|O_{13}^{\delta}|^2+\tfrac{1}{4}Q^2$
& $\tfrac{3}{4}Q^2$
& $\tfrac{1}{6}|O_{13}^{\delta}|^2+\tfrac{1}{4}|W|^2$
& $\tfrac{3}{4}|W|^2$ \\[6pt]
\hline
\hline
\end{tabular}
\end{table}

\begin{figure*}[t]
\centering
\subfloat[NOT gate]{
\includegraphics[width=0.48\textwidth]{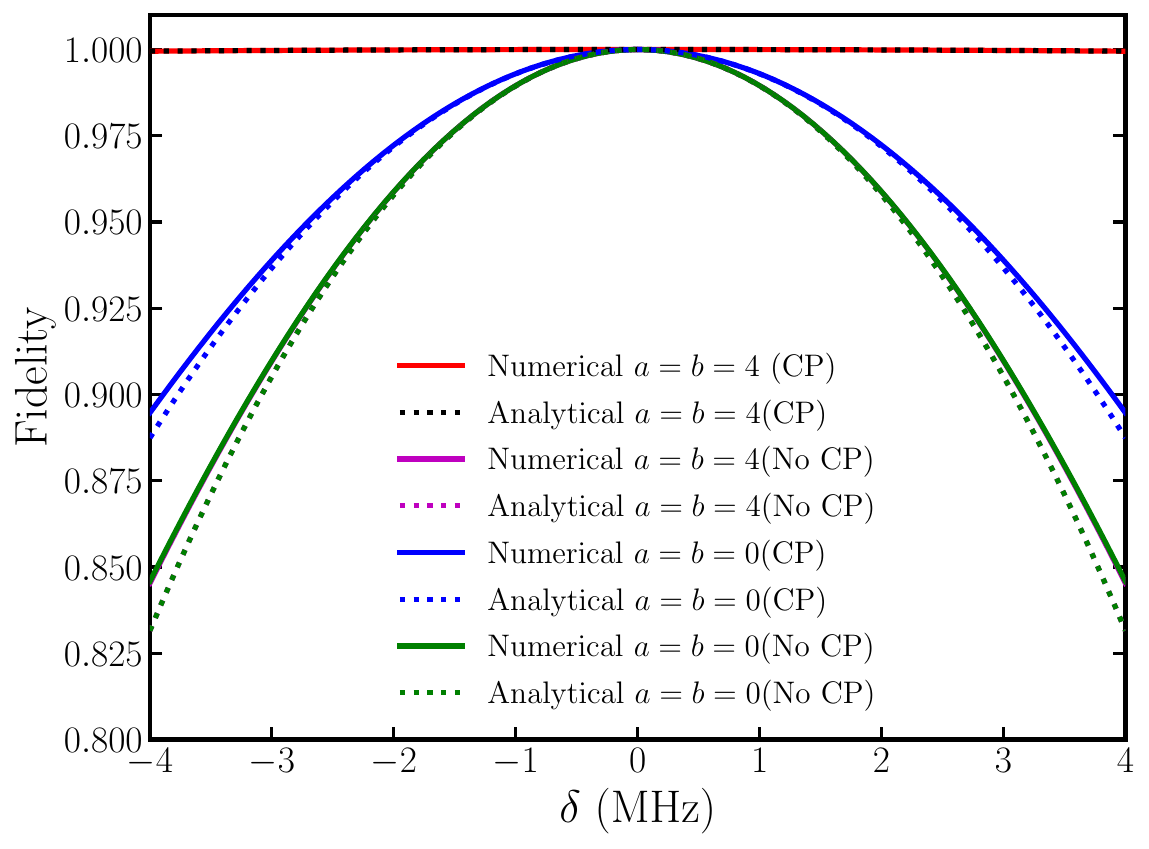}}
\hfill
\subfloat[S gate]{
\includegraphics[width=0.48\textwidth]{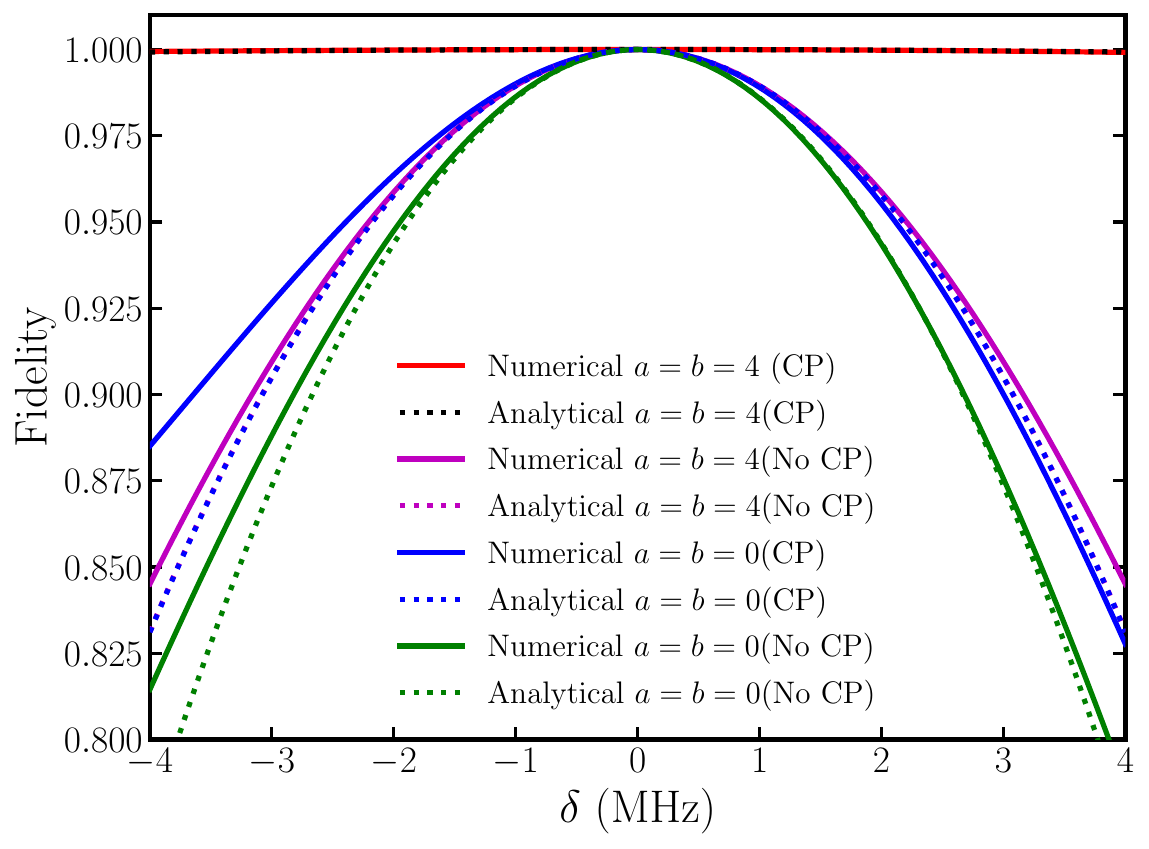}}
\caption{Logical average gate fidelity as a function of the static
detuning error $\delta/2\pi$ for the NOT and S gates.
Solid curves show the numerical results, while dotted curves show the
corresponding second-order perturbative predictions.
The four cases compare the $a=b=0$ and $a=b=4$ trajectories with and
without the compensation pulse.}
\label{fig:fidelity_delta}
\end{figure*}

The analytical expressions reproduce the local quadratic behavior of
the numerical results around $\delta=0$.
At larger detunings, higher-order corrections become visible,
particularly for the $a=b=0$ compensated sequence after the dominant
quadratic error channel has been removed.

\subsection{Combined Error Landscape}
\label{sec:contour}

\begin{figure*}[htbp]
\centering
\subfloat[NOT gate]{
\includegraphics[width=0.48\textwidth]{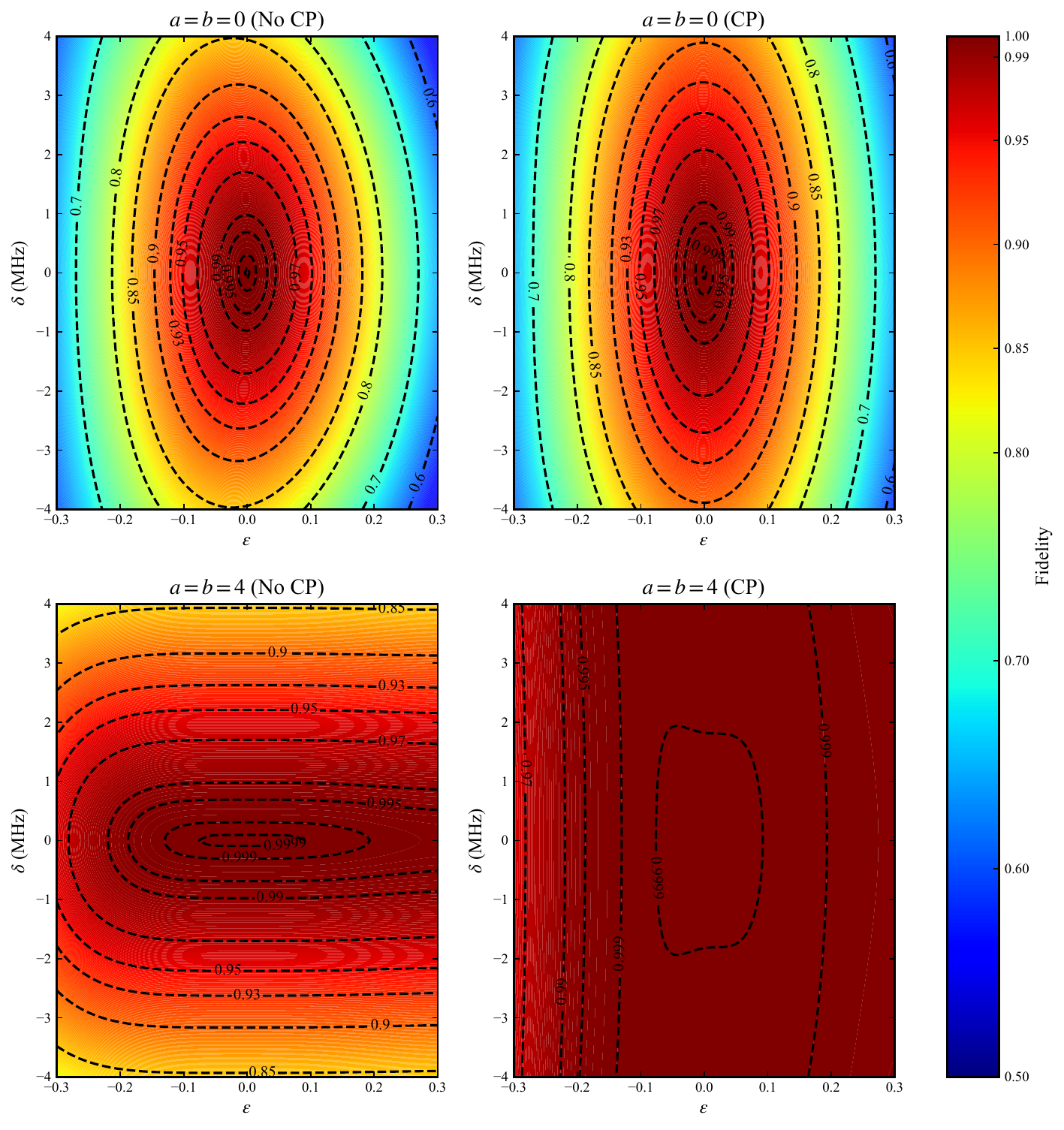}}
\hfill
\subfloat[S gate]{
\includegraphics[width=0.48\textwidth]{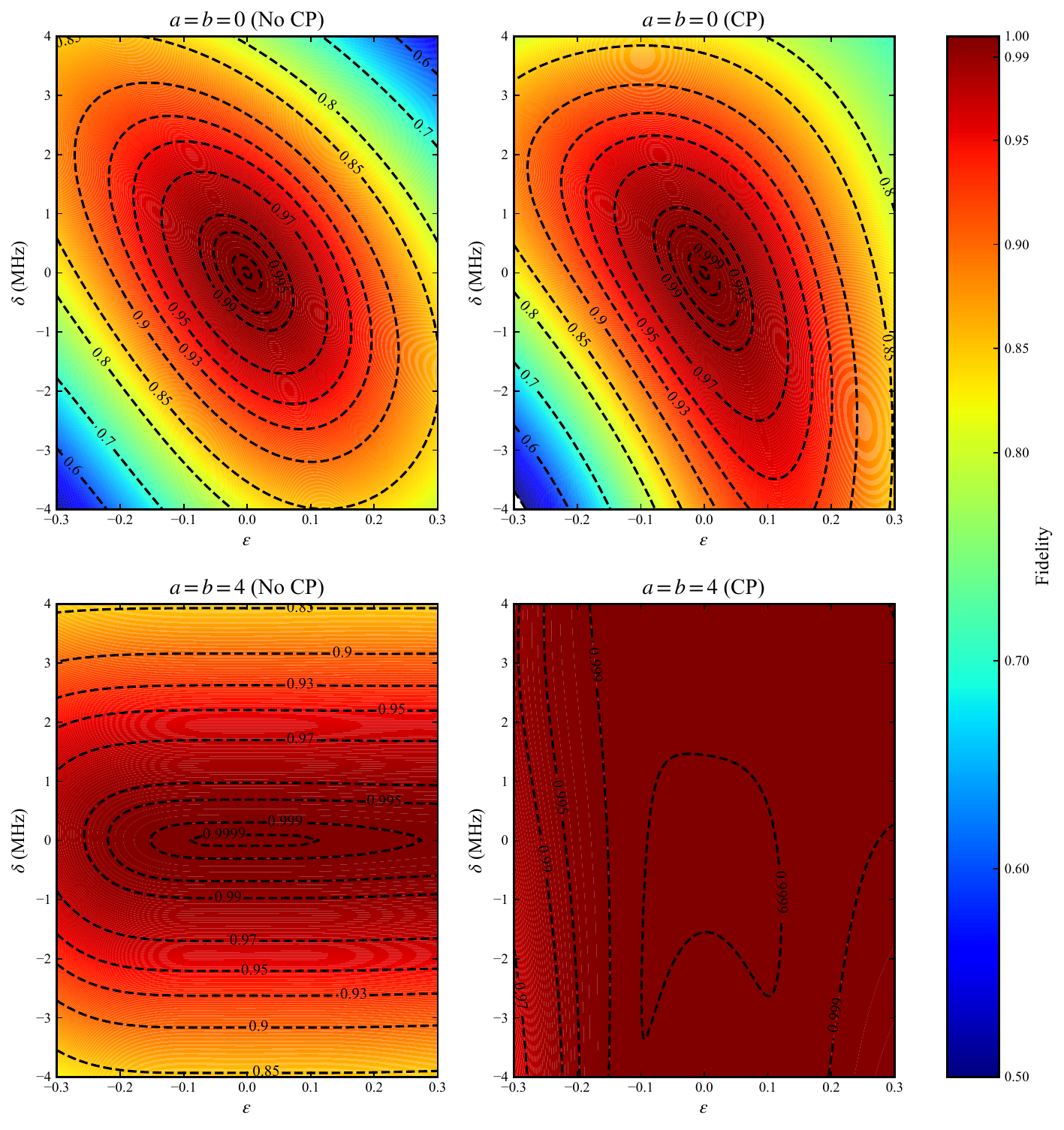}}
\caption{Two-dimensional maps of the logical average gate fidelity
under simultaneous Rabi-amplitude and detuning errors.
The horizontal and vertical axes denote $\epsilon$ and $\delta/2\pi$ (MHz),
respectively. For each gate, the four panels compare the conventional
and optimized pulse trajectories with and without compensation.}
\label{fig:contour_plots}
\end{figure*}

We next consider simultaneous Rabi-amplitude and detuning errors.
Figure~\ref{fig:contour_plots} shows the logical average gate fidelity
over the two-dimensional parameter space $(\epsilon,\delta/2\pi)$ for the
NOT and S gates.

For each gate, four control strategies are compared:

\begin{itemize}

\item Standard NHQC ($a=b=0$, No CP):
the baseline sequence has limited tolerance to both errors.

\item Standard with Compensation ($a=b=0$, CP):
the compensation loop reduces the dominant detuning contribution,
while the sensitivity to Rabi-amplitude errors remains.

\item Optimized Pulse ($a=b=4$, No CP):
pulse shaping substantially increases the robustness along the
Rabi-error direction, but the residual $O_{13}^{\delta}$ contribution
still limits the detuning robustness.

\item Robust Composite Scheme ($a=b=4$, CP):
the optimized pulse and compensation loop act on complementary error
channels and produce the broadest high-fidelity region.

\end{itemize}

\subsection{Quantitative Trade-off Analysis: Systematic Errors vs. Decoherence}
\label{sec:lindblad_analysis}

The compensation pulse doubles the total operation time from $\tau$ to
$2\tau$. Consequently, the reduction of systematic detuning errors must
be balanced against the additional exposure to decoherence. We quantify
this competition using a phenomenological Lindblad master equation,
\begin{equation}
\dot{\rho}
=
-i[H(t),\rho]
+
\sum_k
\left(
L_k\rho L_k^\dagger
-\frac{1}{2}
\left\{
L_k^\dagger L_k,\rho
\right\}
\right).
\label{eq:lindblad_model}
\end{equation}
The auxiliary level $|e\rangle$ is assumed to decay equally to
$|0\rangle$ and $|1\rangle$, with
$L_0=\sqrt{1/(2T_1)}|0\rangle\langle e|$ and
$L_1=\sqrt{1/(2T_1)}|1\rangle\langle e|$.
Pure dephasing is described by
$L_\phi=\sqrt{2\Gamma_\phi}|e\rangle\langle e|$, where
$\Gamma_\phi=1/T_2-1/(2T_1)$.
This model is used as a platform-independent benchmark rather than as
a device-specific microscopic noise model.

As in the closed-system calculations, we evaluate the logical average
gate fidelity using the same six Pauli eigenstates forming a complex
projective two-design. For the Lindblad dynamics, these six input states
are propagated as density matrices, and their corresponding target-state
fidelities are equally averaged. This evaluates the open-system extension
of the Haar-averaged logical-input fidelity, consistently including
decoherence and population leakage; it reduces to
Eq.~(\ref{eq:Favg_definition}) for closed-system evolution.

\begin{figure*}[htbp]
\centering
\subfloat[NOT gate]{
\includegraphics[width=0.48\textwidth]{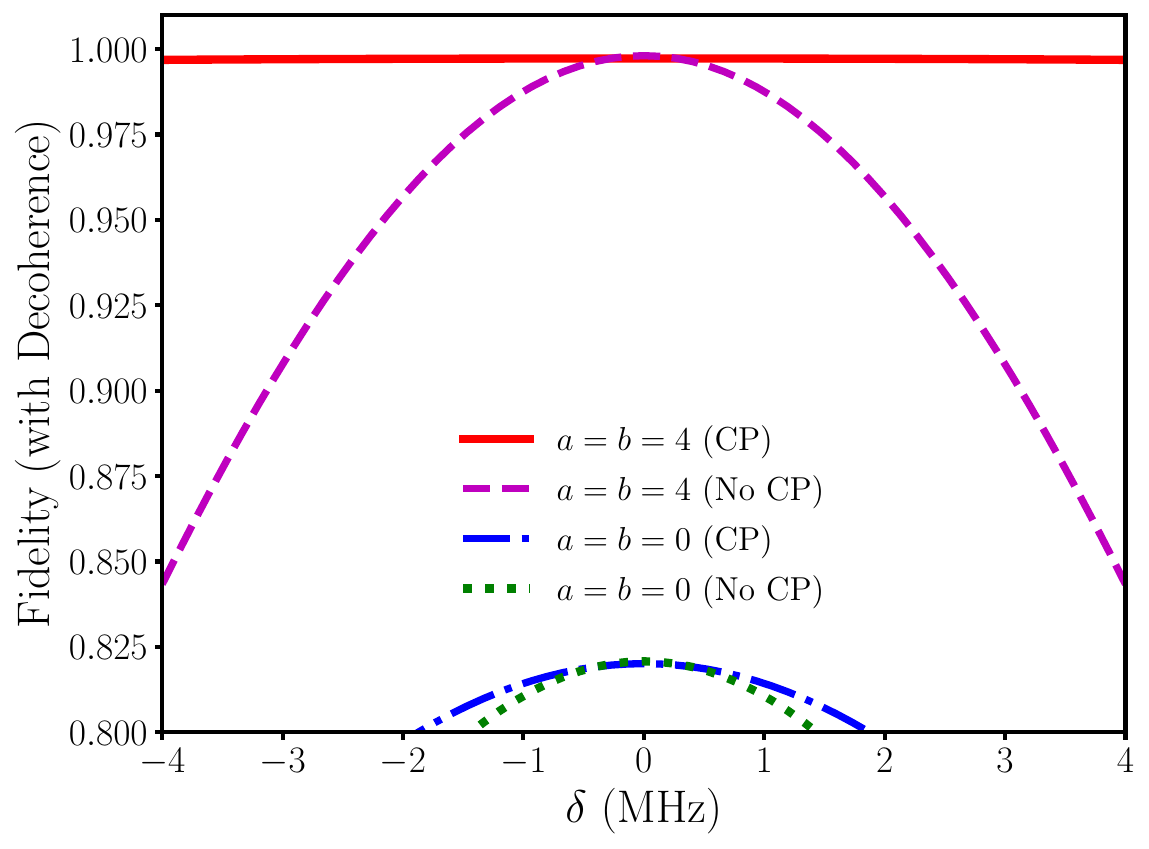}}
\hfill
\subfloat[S gate]{
\includegraphics[width=0.48\textwidth]{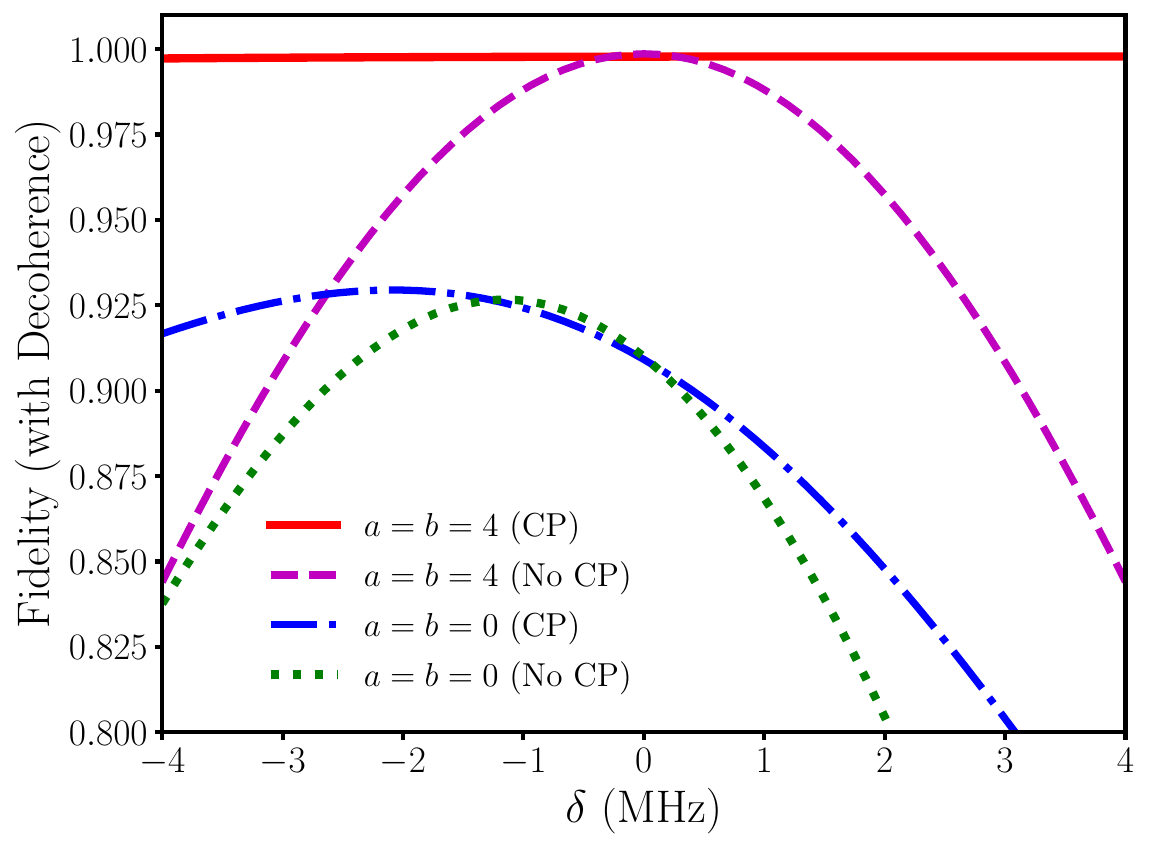}}
\caption{Logical average gate fidelity in the presence of systematic
errors and decoherence as a function of the static detuning
$\delta/2\pi$ for (a) the NOT gate and (b) the S gate.
The parameters are $\epsilon=0.2$ and
$T_1=T_2=30~\mu{\rm s}$.}
\label{fig:lindblad_tradeoff}
\end{figure*}

We first set $T_1=T_2=30~\mu{\rm s}$ and fix the Rabi-amplitude error at
$\epsilon=0.2$. Figure~\ref{fig:lindblad_tradeoff} compares the logical
average gate fidelity of the NOT and S gates for the four control
strategies.

The conventional $a=b=0$ sequence remains strongly sensitive to the
Rabi-amplitude error, irrespective of whether the compensation loop is
added. Pulse shaping with $a=b=4$ suppresses this contribution, leaving
a direct competition between residual detuning error and the additional
decoherence associated with the compensation loop.

At exact resonance, the shorter $a=b=4$ single-loop sequence is slightly
advantageous because the compensation loop provides no detuning-error
benefit while doubling the gate duration. As the detuning increases,
the two optimized strategies cross at approximately
$|\delta|/2\pi\simeq0.28$ MHz for both the NOT and S gates.
Beyond this value, the reduction of the detuning error outweighs the
additional decoherence and the $a=b=4$ composite sequence becomes
advantageous.

At the representative stress-test point
$\epsilon=0.2$ and $\delta/2\pi=2$ MHz, the logical average fidelities
for $a=b=4$ increase from $0.9569$ to $0.9972$ for the NOT gate and
from $0.9574$ to $0.9979$ for the S gate upon adding the compensation
loop. This corresponds to reductions of the gate infidelity by factors
of approximately $15.3$ and $19.9$, respectively.

\begin{figure*}[htbp]
\centering
\subfloat[NOT gate]{
\includegraphics[width=0.48\textwidth]{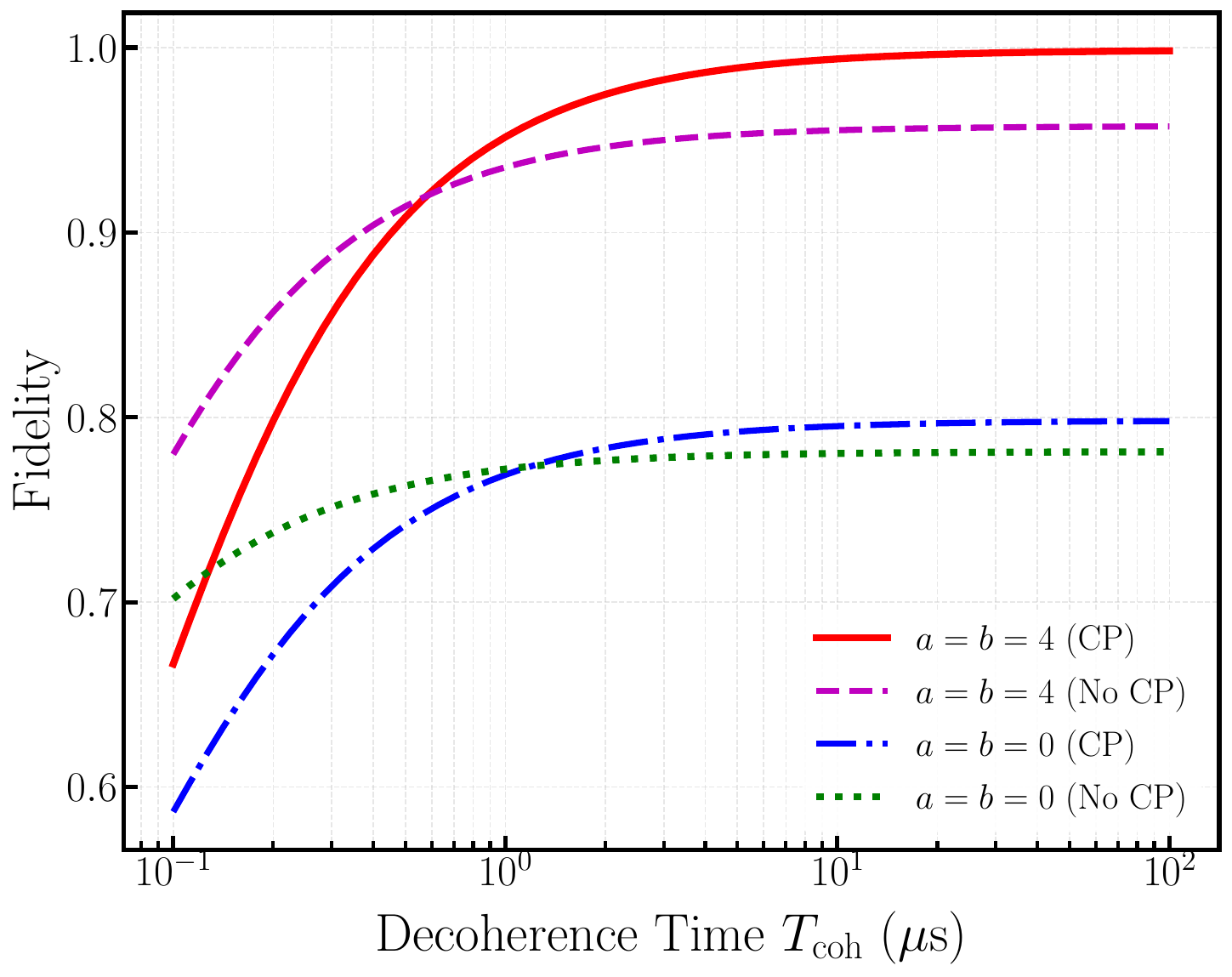}}
\hfill
\subfloat[S gate]{
\includegraphics[width=0.48\textwidth]{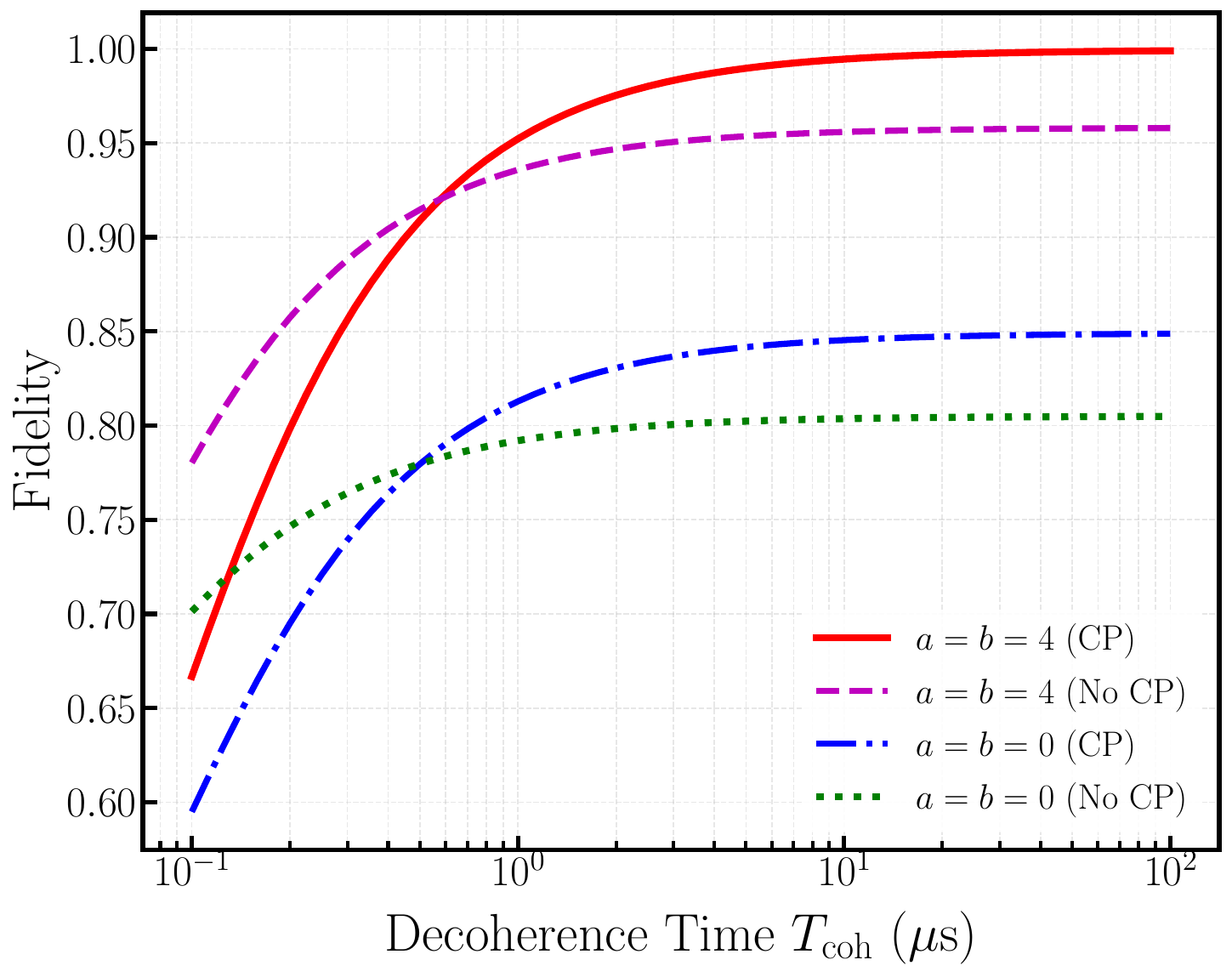}}
\caption{Logical average gate fidelity as a function of the common
coherence time $T_{\rm coh}$ for (a) the NOT gate and (b) the S gate,
with $T_1=T_2=T_{\rm coh}$.
The systematic errors are fixed at $\epsilon=0.2$ and
$\delta/2\pi=2$ MHz. The optimized composite sequence becomes
advantageous above $T_{\rm coh}\simeq0.58~\mu{\rm s}$.}
\label{fig:decoherence_threshold}
\end{figure*}

To further isolate the competition between control-error suppression
and gate duration, we calculate the fidelity as a function of a common
coherence time $T_{\rm coh}$, with
$T_1=T_2=T_{\rm coh}$, while fixing
$\epsilon=0.2$ and $\delta/2\pi=2$ MHz. The results are shown in Fig.~\ref{fig:decoherence_threshold}.

The crossover between the optimized single-loop and composite
strategies occurs at
$T_{\rm coh}^{*}\simeq0.580~\mu{\rm s}$ for the NOT gate and
$T_{\rm coh}^{*}\simeq0.579~\mu{\rm s}$ for the S gate.
Thus, within the present noise model, a common threshold of approximately
$0.58~\mu{\rm s}$ separates the decoherence-dominated regime from the
regime in which systematic-error suppression makes the compensation
loop beneficial. Since the full composite gate lasts only
$2\tau=160$ ns, this crossover corresponds to approximately $3.6$
composite-gate durations.

The optimized composite sequence exceeds $99\%$ logical average
fidelity once $T_{\rm coh}$ reaches approximately
$5.5~\mu{\rm s}$ for the NOT gate and
$5.1~\mu{\rm s}$ for the S gate.
At the representative value $T_{\rm coh}=30~\mu{\rm s}$,
the fidelities reach $0.9972$ and $0.9979$, respectively.
These crossover values are benchmark-dependent and should not be
interpreted as universal hardware thresholds; rather, they quantify
the experimentally relevant operating window for the pulse duration
and systematic-error amplitudes considered here.

\section{Discussion and Conclusion}
\label{sec:conclusion}

We have developed a pulse-engineering strategy for fast and robust
non-adiabatic holonomic quantum gates in a three-level qutrit by
combining inverse engineering with time-dependent perturbation theory.
The optimized $a=b=4$ trajectory
suppresses the leading second-order Rabi-amplitude
contribution through pulse shaping. Static detuning has a different
structure: the dominant $O_{13}^{\delta}$ contribution is associated
with the time-integrated population of the auxiliary excited state
$|e\rangle$ [see Eq.~(\ref{eq:O13delta_estate})] and therefore cannot
be eliminated by single-loop pulse shaping alone. The compensation
loop provides an echo-like sign reversal that
exactly cancels this population-mediated
$O_{13}^{\delta}$ contribution to second order, while the residual
$O_{12}^{\delta}$ channel is further suppressed by the optimized pulse.
This separation of the two error channels provides the physical basis
for the composite protocol.

Numerical simulations for the NOT, Hadamard, S, and T gates show
close agreement with the Haar-averaged second-order
predictions in the perturbative-error regime
(Tables~\ref{Table:infidelity_epsilon} and
\ref{Table:infidelity_delta}). The logical average gate fidelity over
the complete computational subspace confirms that pulse shaping and
compensation act on complementary error channels, producing a
substantially enlarged high-fidelity region under simultaneous
Rabi-amplitude and detuning errors
(Fig.~\ref{fig:contour_plots}).

Experimentally, the protocol requires a controllable three-level
$\Lambda$ manifold driven by two phase-coherent fields, with the pulse
envelopes and common control phase programmed according to
$\Omega_P(t)$, $\Omega_S(t)$, and $\xi(t)$. No additional auxiliary
level or coupling beyond the original $\Lambda$ configuration is
required; the robustness enhancement is implemented through amplitude
and phase shaping of the existing control fields together with the
appended compensation segment. For the parameters considered here, the
optimized single-loop gate has duration $\tau=80$ ns and requires
$\Omega_{\rm max}/2\pi\simeq40.5$ MHz, compared with approximately
$9.82$ MHz for the conventional $a=b=0$ trajectory. The compensated
sequence lasts $2\tau=160$ ns. The principal experimental cost is
therefore a larger peak drive amplitude for pulse shaping and a longer
control window for compensation.

The open-system simulations quantify when this additional control cost
is beneficial. With $T_1=T_2=30~\mu{\rm s}$ and $\epsilon=0.2$, the
optimized composite and single-loop strategies cross at approximately
$|\delta|/2\pi\simeq0.28$ MHz for both the NOT and S gates. At the
representative point $\delta/2\pi=2$ MHz, the composite sequence
increases the logical average fidelity from $0.9569$ to $0.9972$ for
the NOT gate and from $0.9574$ to $0.9979$ for the S gate, corresponding
to reductions of the infidelity by factors of approximately $15.3$ and
$19.9$, respectively. When the common coherence time is varied, the
composite sequence becomes preferable at
$T_{\rm coh}\simeq0.58~\mu{\rm s}$, while $99\%$ fidelity is reached at
approximately $5.5~\mu{\rm s}$ for the NOT gate and
$5.1~\mu{\rm s}$ for the S gate.

These results define a concrete operating regime for the protocol:
control times should remain short compared with the available coherence
time, while quasi-static amplitude-calibration and transition-frequency
errors are sufficiently large that their suppression compensates for
the additional gate duration. The numerical crossover values are not
universal hardware specifications, since the present Lindblad model is
phenomenological and device-specific noise spectra and control-transfer
functions will modify the precise thresholds. They nevertheless provide
quantitative criteria for assessing whether the composite sequence is
advantageous on a given qutrit platform.

The freedom in the inverse-engineered functions $\alpha(t)$ and
$\beta(t)$ permits smooth control envelopes compatible with standard
waveform-generation hardware. Future work will incorporate
device-specific noise spectra and experimentally measured
control-transfer functions, extend the construction to entangling
gates, and combine systematic-error suppression with time-optimal
control~\cite{wang2015quantum,liu2020brachistochrone,xu2018path,
li2020approach,ding2021nonadiabatic,liang2022composite}
to reduce the resource overhead of the composite sequence.
\appendix

\section{Derivation of Second-Order Fidelity Terms}
\label{app:derivation}

This appendix gives the second-order fidelity corrections for the qutrit
system using time-dependent perturbation theory over the gate stage
$t\in[0,\tau]$ and the compensation stage $t\in[\tau,2\tau]$.

\subsection{Fidelity over $t\in[0,\tau]$}

The ideal propagator can be expressed in the time-dependent basis
$\{|d\rangle,|\chi_+(t)\rangle,|\chi_-(t)\rangle\}$ as
\begin{align}
U^{(0)}(t,0)
&=|d\rangle\langle d|
+e^{i\varphi_+(t)}|\chi_+(t)\rangle\langle\chi_+(0)|\nonumber\\
&+e^{i\varphi_-(t)}|\chi_-(t)\rangle\langle\chi_-(0)|.
\label{eq:general_Ut}
\end{align}
Since $\alpha(0)=0$, $|\chi_+(0)\rangle=|b\rangle$ and
$|\chi_-(0)\rangle=-e^{i\beta(0)}|e\rangle$, yielding
\begin{align}
U^{(0)}(t,0)
&=|d\rangle\langle d|
+e^{i\varphi_+(t)}|\chi_+(t)\rangle\langle b|\nonumber\\
&-e^{i[\varphi_-(t)-\beta(0)]}|\chi_-(t)\rangle\langle e|.
\label{eq:Ut_global_phase}
\end{align}

For an arbitrary normalized logical input
$|\psi(0)\rangle=c_d(0)|d\rangle+c_b(0)|b\rangle$, we choose the
orthonormal basis
\begin{align}
|\psi_1^{(0)}(t)\rangle
&=U^{(0)}(t,0)[c_d(0)|d\rangle+c_b(0)|b\rangle],
\label{eq:complete_basisA1}\\
|\psi_2^{(0)}(t)\rangle
&=U^{(0)}(t,0)|e\rangle,
\label{eq:complete_basisA2}\\
|\psi_3^{(0)}(t)\rangle
&=U^{(0)}(t,0)[-c_b^*(0)|d\rangle+c_d^*(0)|b\rangle].
\label{eq:complete_basisA3}
\end{align}
Its orthonormality and completeness are preserved by the unitarity of
$U^{(0)}(t,0)$.

The perturbation matrix elements are
$H_{1m}^{'}(t)=
\langle\psi_1^{(0)}(t)|H^{'}(t)|\psi_m^{(0)}(t)\rangle$,
with
\begin{equation}
H^{'}(t)=\delta|e\rangle\langle e|
+\epsilon\Omega(t)
\left[
e^{-i\xi(t)}|b\rangle\langle e|
+\mathrm{h.c.}
\right].
\label{eq:Hprime_app}
\end{equation}

The two relevant off-diagonal elements are
\begin{align}
H_{12}^{'}(t)
&=
\frac{1}{2}e^{-i\beta(0)}c_b^*(0)
e^{-i[\varphi_+(t)-\varphi_-(t)]}
\left\{
\delta\sin\alpha(t)
-\epsilon\left[
\dot{\beta}(t)\sin\alpha(t)
+i\dot{\alpha}(t)
\right]
\right\},
\label{eq:H12_prime_app}\\
H_{13}^{'}(t)
&=
c_d^*(0)c_b^*(0)
\left[
\delta\sin^2\frac{\alpha(t)}{2}
-\frac{\epsilon}{2}\dot{\beta}(t)
\sin\alpha(t)\tan\alpha(t)
\right].
\label{eq:H13_general_prime_app}
\end{align}

Setting $\epsilon=0$ gives
\begin{equation}
H_{13}^{'}(t)
=
\delta c_d^*(0)c_b^*(0)
\sin^2\frac{\alpha(t)}{2}.
\label{eq:H13_prime_app}
\end{equation}

\subsection{Fidelity over $t\in[\tau,2\tau]$}

The compensation stage uses
$\tilde{\theta}=\pi-\theta$ and $\tilde{\phi}=\pi+\phi$.
The same orthonormal basis is propagated continuously through the second
loop according to
\begin{equation}
|\tilde{\psi}_n^{(0)}(t)\rangle
=\tilde U^{(0)}(t,\tau)U^{(0)}(\tau,0)|\psi_n^{(0)}(0)\rangle,
\qquad n=1,2,3,
\label{eq:complete_basis_second_stage}
\end{equation}
so that
$|\tilde{\psi}_n^{(0)}(\tau)\rangle=|\psi_n^{(0)}(\tau)\rangle$.
For pure detuning error, the relevant matrix elements are
\begin{align}
\tilde H_{12}^{'}(t)
&=\frac{\delta}{2}e^{-i\beta(\tau)}c_d^*(0)
e^{-i[\tilde{\varphi}_+(t)-\tilde{\varphi}_-(t)]}\sin\alpha(t),
\label{eq:H12_tilde_prime_app}\\
\tilde H_{13}^{'}(t)
&=-\delta c_d^*(0)c_b^*(0)\sin^2\frac{\alpha(t)}{2}.
\label{eq:H13_tilde_prime_app}
\end{align}
The compensation transformation therefore gives
\begin{equation}
\tilde H_{13}^{'}(t)=-H_{13}^{'}(t-\tau),
\label{eq:H13_sign_reversal_app}
\end{equation}
so the $m=3$ contribution cancels in the second-order fidelity.

The remaining $m=2$ contribution takes the general two-stage form
\begin{align}
P(2\tau)
&\simeq1-\left|\int_0^\tau dt\,H_{12}^{'}(t)\right|^2
-\left|\int_\tau^{2\tau}dt\,\tilde H_{12}^{'}(t)\right|^2 \notag\\
&\quad-\int_0^\tau dt_1\int_\tau^{2\tau}dt_2
\left[H_{21}^{'}(t_1)\tilde H_{12}^{'}(t_2)+\mathrm{h.c.}\right].
\label{eq:P2tau_general_residual}
\end{align}
Defining
\begin{align}
\mathcal I_1
&=e^{-i\beta(0)}\int_0^\tau dt\,
e^{-i[\varphi_+(t)-\varphi_-(t)]}\sin\alpha(t),
\label{eq:app_I1}\\
\mathcal I_2
&=e^{-i\beta(\tau)}\int_\tau^{2\tau}dt\,
e^{-i[\tilde{\varphi}_+(t)-\tilde{\varphi}_-(t)]}\sin\alpha(t),
\label{eq:app_I2}
\end{align}
the state-resolved fidelity becomes
\begin{align}
P(2\tau)
&\simeq1-\frac{\delta^2}{4}\Big[
|c_b(0)|^2|\mathcal I_1|^2+|c_d(0)|^2|\mathcal I_2|^2 \notag\\
&\qquad+2\,{\rm Re}\!\left(
c_d^*(0)c_b(0)\mathcal I_1^*\mathcal I_2\right)\Big].
\label{eq:app_P2tau_3level}
\end{align}
Thus, the compensation loop exactly cancels the population-mediated
$O_{13}^{\delta}$ contribution to second order, while the residual
detuning sensitivity is determined by the $O_{12}^{\delta}$ channel.

\section{Haar-Averaged Logical Gate Fidelity}
\label{app:haar_average_fidelity}

To compare the perturbative state fidelity directly with the numerical
logical average gate fidelity, we average over the complete logical
input space. We parameterize
\begin{equation}
|\psi(0)\rangle=\cos\theta_0|0\rangle+e^{i\phi_0}\sin\theta_0|1\rangle,
\label{eq:app_haar_input}
\end{equation}
where $0\leq\theta_0\leq\pi/2$ and $0\leq\phi_0<2\pi$.

The normalized Haar measure is
\begin{equation}
d\mu_{\rm H}=\frac{1}{2\pi}\sin(2\theta_0)\,d\theta_0\,d\phi_0.
\label{eq:app_haar_measure}
\end{equation}
Using Eqs.~(\ref{eq:dark}) and (\ref{eq:bright}),
\begin{equation}
|\psi(0)\rangle=c_d(0)|d\rangle+c_b(0)|b\rangle,
\label{eq:app_haar_db_state}
\end{equation}
with
\begin{align}
c_d(0)&=\cos\theta_0\cos\frac{\theta}{2}
+\sin\theta_0\sin\frac{\theta}{2}e^{i(\phi_0-\phi)},
\label{eq:app_cd}\\
c_b(0)&=\cos\theta_0\sin\frac{\theta}{2}
-\sin\theta_0\cos\frac{\theta}{2}e^{i(\phi_0-\phi)}.
\label{eq:app_cb}
\end{align}

Since this basis transformation is unitary, the Haar measure is
unchanged, and the required moments are
\begin{equation}
\langle|c_d|^2\rangle_{\rm H}=\langle|c_b|^2\rangle_{\rm H}=\tfrac12,\quad
\langle|c_d|^2|c_b|^2\rangle_{\rm H}=\tfrac16,\quad
\langle c_d^*c_b\rangle_{\rm H}=0.
\label{eq:app_haar_moments}
\end{equation}

For an evolution of duration $T$, where $T=\tau$ for a single loop and
$T=2\tau$ for a compensated sequence, the projected logical operation
and its average gate fidelity are
\begin{equation}
M=P_{\rm L}U(T)P_{\rm L},
\label{eq:app_haar_M}
\end{equation}
\begin{align}
F_{\rm avg}
&=\int d\mu_{\rm H}(\psi)
\left|\langle\psi|U_{\rm tg}^{\dagger}M|\psi\rangle\right|^2 \notag\\
&=\frac{{\rm Tr}(M^\dagger M)+|{\rm Tr}(U_{\rm tg}^{\dagger}M)|^2}{d(d+1)},
\qquad d=2.
\label{eq:app_haar_Favg}
\end{align}
The numerical calculation uses the six Pauli eigenstates
\begin{equation}
|0\rangle,\quad |1\rangle,\quad
|\pm\rangle=\frac{|0\rangle\pm|1\rangle}{\sqrt2},\quad
|\pm i\rangle=\frac{|0\rangle\pm i|1\rangle}{\sqrt2}.
\label{eq:app_six_states}
\end{equation}

These states form a complex projective two-design, so their equally
weighted average reproduces the Haar average exactly for the
single-qubit fidelity considered here.

The state-resolved second-order result is
\begin{equation}
P_\psi(\tau)\simeq1-\frac14|c_b|^2|\delta O_{12}^{\delta}+\epsilon O_{12}^{\epsilon}|^2
-|c_d|^2|c_b|^2|\delta O_{13}^{\delta}+\epsilon O_{13}^{\epsilon}|^2.
\label{eq:app_haar_general_state}
\end{equation}
Averaging with Eq.~(\ref{eq:app_haar_moments}) gives
\begin{equation}
F_{\rm avg}^{(2)}(\tau)\simeq1-\frac18|\delta O_{12}^{\delta}+\epsilon O_{12}^{\epsilon}|^2
-\frac16|\delta O_{13}^{\delta}+\epsilon O_{13}^{\epsilon}|^2.
\label{eq:app_haar_general}
\end{equation}
Setting $\delta=0$, the conventional $a=b=0$ trajectory satisfies
\begin{equation}
O_{13}^{\epsilon}=0,\quad
O_{12}^{\epsilon}=-i\pi(1-e^{-i\gamma}),\quad
|O_{12}^{\epsilon}|^2=2\pi^2(1-\cos\gamma).
\label{eq:app_haar_rabi_O}
\end{equation}

Hence,
\begin{equation}
F_{\mathrm{avg},\epsilon}^{(2)}
=1-\frac{\pi^2}{4}(1-\cos\gamma)\epsilon^2.
\label{eq:app_haar_rabi_result}
\end{equation}
This gives the coefficients in Table~\ref{Table:infidelity_epsilon}.
In contrast, $a=b=4$ gives
$O_{12}^{\epsilon}=O_{13}^{\epsilon}=0$, so that
\begin{equation}
1-F_{\mathrm{avg},\epsilon}=\mathcal O(\epsilon^4).
\label{eq:app_haar_rabi_ab4}
\end{equation}

Turning to static detuning, setting $\epsilon=0$ gives
\begin{equation}
F_{\mathrm{avg},\delta}^{\mathrm{NoCP}}
=1-\frac{\delta^2}{8}|O_{12}^{\delta}|^2
-\frac{\delta^2}{6}|O_{13}^{\delta}|^2.
\label{eq:app_haar_detuning_start}
\end{equation}
For the symmetric trajectories,
\begin{equation}
O_{12}^{\delta}=A(1+e^{-i\gamma}),\qquad
|O_{12}^{\delta}|^2=2|A|^2(1+\cos\gamma),
\label{eq:app_haar_detuning_A}
\end{equation}
where $A=Q$ for $a=b=0$ and $A=W$ for $a=b=4$.

The single-loop result therefore becomes
\begin{equation}
F_{\mathrm{avg},\delta}^{\mathrm{NoCP}}
=1-\delta^2\left[
\frac{|O_{13}^{\delta}|^2}{6}
+\frac{|A|^2}{4}(1+\cos\gamma)\right].
\label{eq:app_haar_detuning_NoCP}
\end{equation}
With compensation, Eqs.~(\ref{eq:app_P2tau_3level}) and
(\ref{eq:app_haar_moments}) show that the input-state-dependent
interference term vanishes under the Haar average, yielding
\begin{equation}
F_{\mathrm{avg},\delta}^{\mathrm{CP}}
=1-\frac{\delta^2}{8}\left(|\mathcal I_1|^2+|\mathcal I_2|^2\right).
\label{eq:app_haar_CP_intermediate}
\end{equation}
The pulse symmetry gives
\begin{equation}
|\mathcal I_1|^2=2|A|^2(1+\cos\gamma),\qquad
|\mathcal I_2|^2=4|A|^2.
\label{eq:app_haar_CP_integrals}
\end{equation}
Thus,
\begin{equation}
F_{\mathrm{avg},\delta}^{\mathrm{CP}}
=1-\frac{\delta^2|A|^2}{4}(3+\cos\gamma).
\label{eq:app_haar_detuning_CP}
\end{equation}
Equations~(\ref{eq:app_haar_detuning_NoCP}) and
(\ref{eq:app_haar_detuning_CP}) give the coefficients in
Table~\ref{Table:infidelity_delta} and the analytical curves in
Fig.~\ref{fig:fidelity_delta}.

\section*{ACKNOWLEDGEMENTS} 

The authors gratefully acknowledge Prof. Xi Chen and  Shi-Lei Su for helpful discussions. This work was supported by the China Scholarship Council (CSC), the Independent Research Project of the Key Laboratory of Advanced Optical Manufacturing Technologies of Jiangsu Province (Grant No. ZZ2109), the Natural Science Foundation of Jiangsu Province (Grant Nos. BG2025017 and BK20250404), the Youth Science and Technology Talent Support Project of Jiangsu Province (Grant Nos. JSTJ-2025-600 and JSTJ-2025-829), and the Frontier Technology Research Program of Suzhou (Grant No. SYG202322).

\bibliographystyle{apsrev4-2}
\bibliography{Robust_three_level}

\end{document}